\documentclass[reprint,aps,prl,floatfix,superscriptaddress]{revtex4-1}
\usepackage{}
\usepackage{amssymb,amsmath,amsfonts}
\usepackage{bm}
\usepackage{color}
\usepackage{graphicx}

\usepackage{hyperref}
\usepackage{multirow}

\DeclareMathSizes{10}{10}{6}{4}
\let\sss=\scriptscriptstyle

\begin{document}
\title{Observation of Four-body Ring-exchange Interactions \\and Anyonic Fractional Statistics 
}

\author{Han-Ning Dai$^{1,2,3,*}$, Bing Yang$^{1,2,3,*}$, Andreas Reingruber$^{2,5}$, Hui Sun$^{1,3}$, Xiao-Fan Xu$^{2}$, Yu-Ao Chen$^{1,3,4}$, Zhen-Sheng Yuan$^{1,2,3,4}$, Jian-Wei Pan$^{1,2,3,4}$\\ \vspace{1mm}
\textit{\small $^1$Hefei National Laboratory for Physical Sciences at Microscale and Department of Modern \\ \vspace{-1mm}Physics, University of Science and Technology of China, Hefei, Anhui 230026, China.}\\
\textit{\small $^2$Physikalisches Institut, Ruprecht-Karls-Universit\"{a}t Heidelberg, Im Neuenheimer Feld 226, 69120 Heidelberg, Germany}\\
\textit{\small $^3$CAS Centre for Excellence and Synergetic Innovation Centre in Quantum Information \\ \vspace{-1mm}and Quantum Physics, University of Science and Technology of China, Hefei, Anhui 230026, China}\\
\textit{\small $^4$CAS-Alibaba Quantum Computing Laboratory, Shanghai 201315, China}\\
\textit{\small $^5$Department of Physics and Research Center OPTIMAS, University of Kaiserslautern,\\ \vspace{-1mm} Erwin-Schroedinger-Strasse, Building 46, 67663 Kaiserslautern, Germany}\\
\textit{\small $^*$These authors contributed equally to this work.}
}

\date{\today}

\begin{abstract}
Ring exchange is an elementary interaction for modeling unconventional topological matters which hold promise for efficient quantum information processing. We report the observation of four-body ring-exchange interactions and the topological properties of anyonic excitations within an ultracold atom system. A minimum toric code Hamiltonian in which the ring exchange is the dominant term, was implemented by engineering a Hubbard Hamiltonian that describes atomic spins in disconnected plaquette arrays formed by two orthogonal superlattices. The ring-exchange interactions were resolved from the dynamical evolutions in the spin orders, matching well with the predicted energy gaps between two anyonic excitations of the spin system. A braiding operation was applied to the spins in the plaquettes and an induced phase $1.00(3)\pi$ in the four-spin state was observed, confirming $\frac{1}{2}$-anynoic statistics. This work represents an essential step towards studying topological matters with many-body systems and the applications in quantum computation and simulation.
\end{abstract}
\maketitle

\noindent

Exploiting the laws of quantum mechanics, quantum information processing can be exponentially faster than the classical counterpart\cite{nielson2000quantum}. To make this technology a reality, scientists have to solve the crucial problem of decoherence and systematic errors in real quantum systems, which is very difficult due to the request of an extremely small error threshold to enable error corrections \cite{steane1996error,knill2005quantum}. A very encouraging solution to this problem is the Kitaev model \cite{kitaev2003fault} of fault-tolerant quantum computation by anyons, a sort of topological quasiparticles being neither bosons nor fermions \cite{wilczek1982quantum}. In this model, anyons are exploited to encode and manipulate information in a manner which is resistant to errors, the so-called topological protection.
Unfortunately, except that signatures of anyonic statistics emerged in the fractional quantum Hall systems \cite{nayak2008non,willetPRL2013}, there has been no conclusive observation of anyons in any existing matters. A proposal suggests to solely mimic anyonic statistics with non-interacting qubits \cite{han2007scheme} and experimental demonstrations were achieved with entangled photons \cite{lu2009demonstrating,pachos2009revealing} and ions \cite{barreiro2011open}. However, because the background interacting Hamiltonian does not exist in such systems, it is not possible to define anyonic excitations \cite{jiang2008anyonic}. Therefore, the observation of anyons remains challenging.

\begin{figure}[!t]
\centerline{\includegraphics[width=\linewidth]{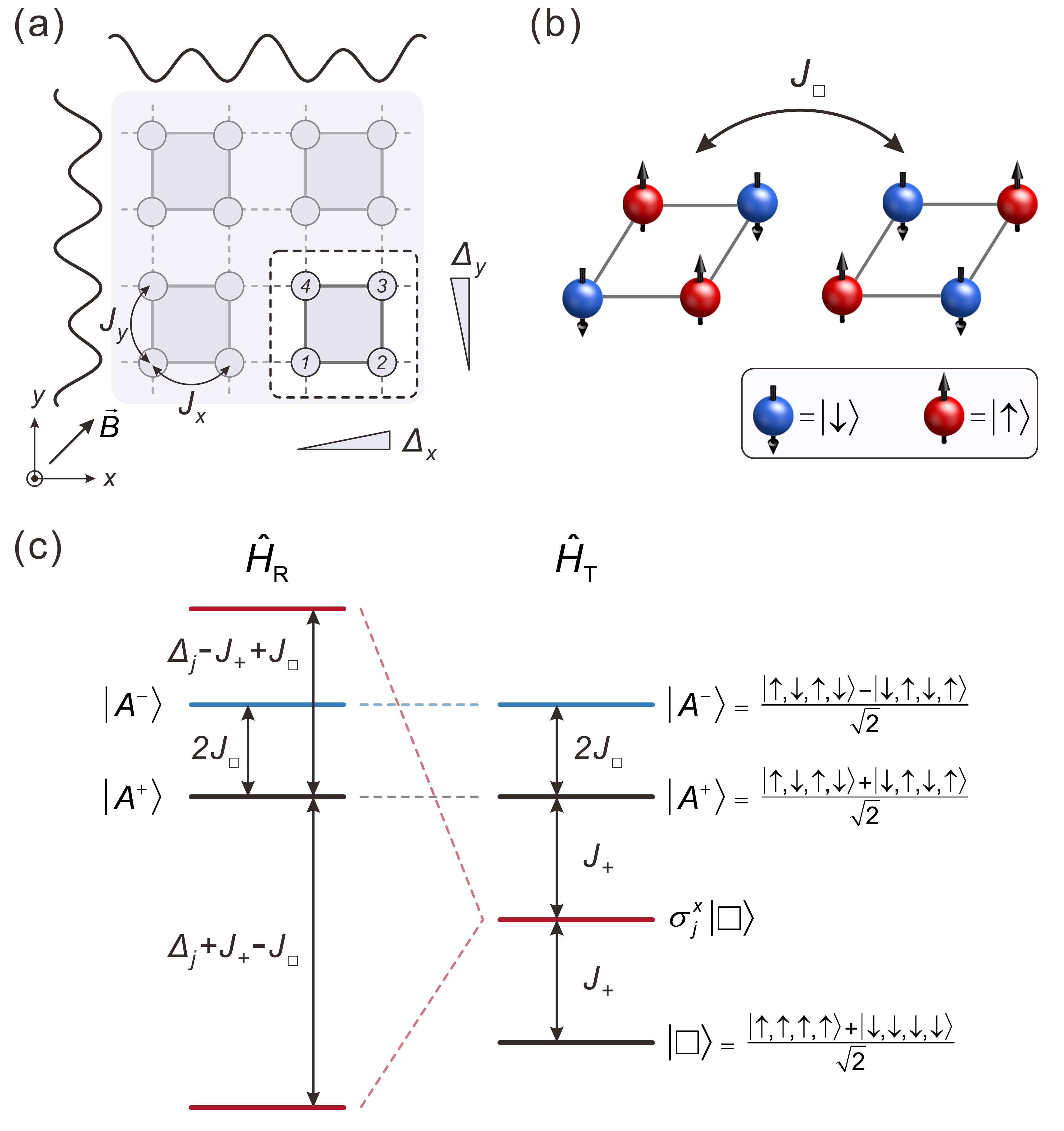}}
\caption{Experimental scheme and the ring-exchange in disconnected four-site plaquettes. \textbf{(a)} Optical plaquettes with effective magnetic gradients are created by two orthogonal spin-dependent superlattices, and the sites of each plaquette are enumerated in a counter-clockwise fashion; \textbf{(b)} The ring-exchange driven oscillations take place between the two antiferromagnetically ordered states \hbox{$|\!\downarrow,\uparrow,\downarrow,\uparrow\rangle$} and \hbox{$|\!\downarrow,\uparrow,\downarrow,\uparrow\rangle$}; \textbf{(c)} Modeling a minimal instance of the Kitaev's toric code $\hat{H}_{\sss \mathrm{T}}$ with the ring-exchange dominating Hubbard Hamiltonian $\hat{H}_{\sss \mathrm{R}}$, in which the quasiparticles are well defined.}
\end{figure}

To construct the appropriate Hamiltonian for studying anyons, a practical scheme \cite{paredes2008minimum} was proposed to create artificial topological matters by manipulating ring-exchange interactions \cite{buchler2005atomic} among ultracold atoms in optical lattices \cite{jaksch1998cold,bloch2008many}. Although a large category of many-body models \cite{greiner2002quantum,simon2011quantum,trotzky2008time,brownSci2015,jotzu2014experimental} have been realized with optical lattices, implementing the ring-exchange Hamiltonian is notoriously difficult due to its nature of the fourth-order spin interaction, which is greatly suppressed compared to the lower order processes, such as superexchange interactions \cite{trotzky2008time,brownSci2015}. So, generation and observation of the ring-exchange interactions and the correlated anyonic excitations become the urgent needs for further studying their physical properties and various attractive applications.

In this work, we implement a four-body ring-exchange Hamiltonian \cite{paredes2008minimum}, the minimum instance of the Kitaev's toric code model \cite{kitaev2003fault}, by carefully suppressing superexchange interactions between ultracold atoms in disconnected optical plaquettes. The ring-exchange interactions are observed by measuring the dynamical evolution of atomic spin configurations in the plaquettes. Moreover, the fractional statistics of Abelian anyons \cite{kitaev2003fault} is demonstrated by a braiding operation and an interference process. As massive optical plaquettes are manipulated in parallel, appropriately connecting the plaquettes offers a great chance to study anyons and topological matters \cite{WenBookQFT} in a large many-body system.

The system under consideration is a four-site plaquette singly occupied with four bosonic atoms in two internal states \hbox{$|\!\downarrow\rangle$} and \hbox{$|\!\uparrow\rangle$} (Fig.1a). It can be well described by a two-species single-band Bose-Hubbard model (BHM), characterized by the tunneling matrix element $J$, an on-site interaction $U$ and a spin-dependent intra-plaquette gradient ${\it \Delta_\mathrm{x(y)}}$. In the regime of strong interactions \hbox{$J \!\ll\! U$} and \hbox{$4J^2/U\!\ll\!{\it \Delta_\mathrm{x(y)}}$}, both the bare tunneling and the superexchange interactions are suppressed, while the fourth-order ring-exchange interactions become dominant. Hence the four-site Hubbard Hamiltonian is reduced to (see supplementary materials and Ref. \cite{paredes2008minimum})
\begin{align}\label{eqnring}
\hat{H}_{\sss \mathrm{R}} =& -J_{\sss \square}(\hat{S}_1^+ \hat{S}_2^- \hat{S}_3^+ \hat{S}_4^- + \mathrm{H.c.}) - J_{\sss +} \sum_{\langle j,k\rangle} \hat{S}_j^z \hat{S}_k^z \nonumber\\&
 +\sum_j {\it \Delta}_j \hat{S}_j^z,
\end{align}
in which \hbox{$J_{\sss \square}\!\approx\!40J^4/U^3$} describes the effective ring-exchange interaction arising from a fourth order tunnelling process, \hbox{$J_{\sss +}\!\approx\!4J^2/U$} denotes a nearest-neighbor Ising-type interaction, and ${\it \Delta}_j$ are the spin-dependent potential biases on site $j$. Therefore the intra-plaquette gradients along the two directions are defined as \hbox{${\it \Delta}_{\mathrm{x}}\!=\!{\it\Delta}_2\!-\!{\it\Delta}_1$} and \hbox{${\it \Delta}_{\mathrm{y}}\!=\!{\it\Delta}_4\!-\!{\it\Delta}_1$}. The corresponding spin operators are defined through the bosonic creation and annihilation operators $\hat{a}^\dagger_{\sigma,j}$ and $\hat{a}_{\sigma,j}$ as \hbox{$\hat{S}^{+}_j\!=\!\hat{a}^\dagger_{\uparrow,j}\hat{a}_{\downarrow,j}$}, \hbox{$\hat{S}^{-}_j\!=\!\hat{a}^\dagger_{\downarrow,j}\hat{a}_{\uparrow,j}$} and \hbox{$\hat{S}^z_j \!=\! (\hat{a}^\dagger_{\uparrow,j}\hat{a}_{\uparrow,j}\!-\!\hat{a}^\dagger_{\downarrow,j}\hat{a}_{\downarrow,j})/2$}, with the two spin states \hbox{$\sigma =\ \uparrow,\downarrow$}. Due to the spatial symmetry of the Hamiltonian, the ring-exchange process can only take place between the two antiferromagnetically ordered states: \hbox{$|\!\downarrow,\uparrow,\downarrow,\uparrow\rangle$} and \hbox{$|\!\uparrow,\downarrow,\uparrow,\downarrow\rangle$}, where the commas separate the occupations of the four sites. Therefore, a system initialized to \hbox{$|\!\downarrow,\uparrow,\downarrow,\uparrow\rangle$} will evolve to \hbox{$|\!\uparrow,\downarrow,\uparrow,\downarrow\rangle$} under the ring-exchange dominating Hamiltonian, and vice versa (Fig.1b). This dynamical process directly reflects the energy gap $2J_{\sss \square}$ between the two eigenstates of $\hat{H}_\mathrm{\sss R}$ (Fig.1c), \hbox{$| A^{\sss +}\rangle \!=\! (|\!\uparrow,\downarrow,\uparrow,\downarrow\rangle\! +\! |\!\downarrow,\uparrow,\downarrow,\uparrow\rangle)/\sqrt{2}$} and  \hbox{$| A^{\sss -}\rangle \!=\! (|\!\uparrow,\downarrow,\uparrow,\downarrow\rangle\! -\! |\!\downarrow,\uparrow,\downarrow,\uparrow\rangle)/\sqrt{2}$}. We use the spin imbalance \hbox{$N_z\!=\! \langle \hat{S}^z_1\!-\!\hat{S}^z_2\!+\!\hat{S}^z_3\!-\!\hat{S}^z_4 \rangle/2$} to characterize the ring-exchange dynamics. Here, $\langle \hat{S}^z_j\rangle$ denote the corresponding quantum mechanical expectation values of $\hat{S}^z$ on site $j$.

\begin{figure*}[!t]
\centerline{\includegraphics[width=13.0cm]{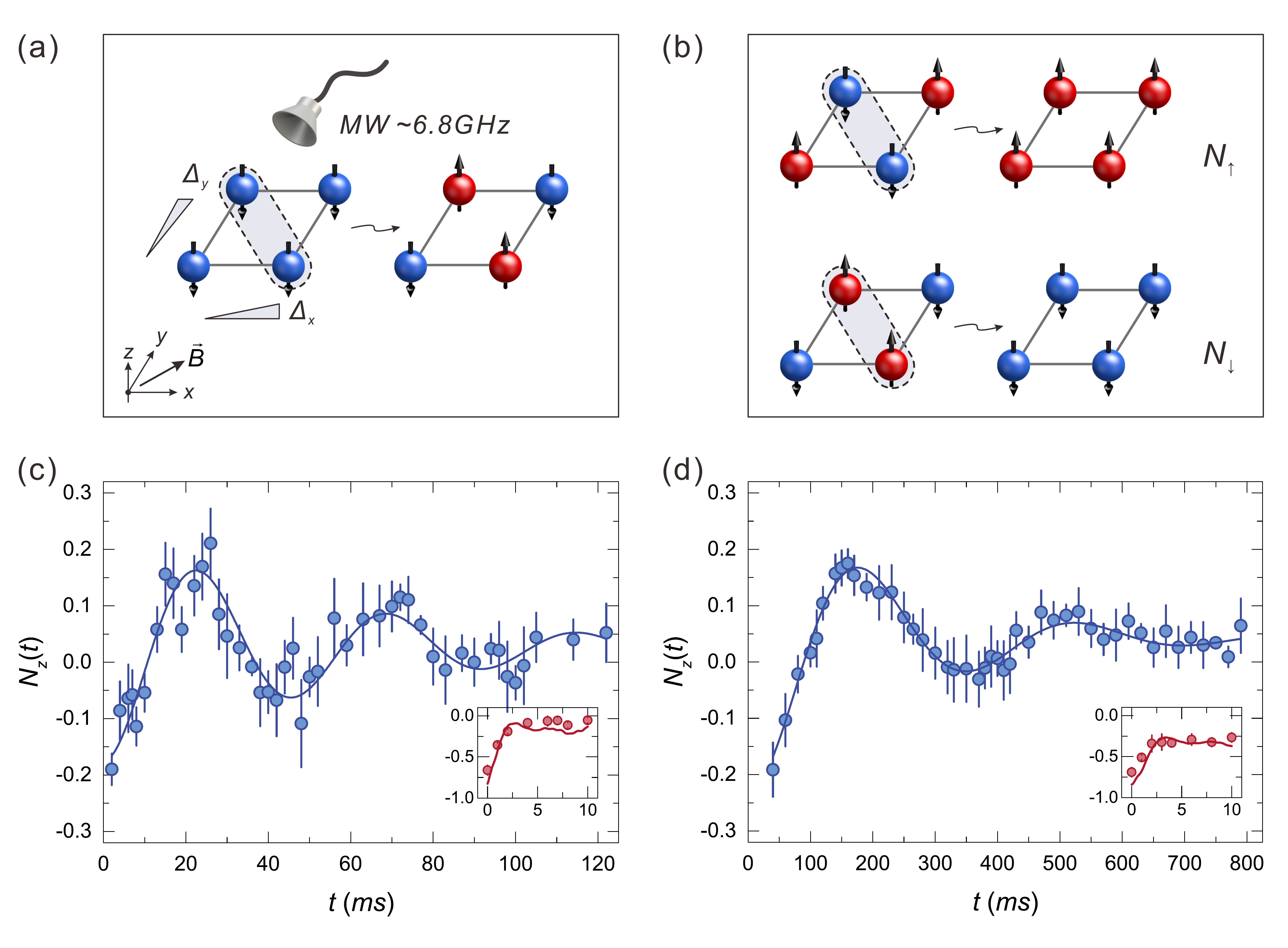}}
\caption{Observation of four-body ring-exchange interactions. \textbf{(a)} A spin-dependent potential lifts the degeneracy of spins on the four sites. The system is initialized from \hbox{$|\!\downarrow,\downarrow,\downarrow,\downarrow\rangle$} to \hbox{$|\!\downarrow,\uparrow,\downarrow,\uparrow\rangle$} by selectively addressing site 2 and 4 with a MW field. \textbf{(b)} Spin configurations are read out by first addressing and flipping the sites 2 and 4, and then imaging the two spin components by taking two in-situ absorption images $N_{\uparrow}(\mathrm{x},\mathrm{y})$ and $N_{\downarrow}(\mathrm{x},\mathrm{y})$ in sequence.  Measured spin dynamics of $N_z(t)$ (blue circles) in isolated plaquettes in the presence of an effective magnetic gradient are shown for two different settings: \textbf{(c)} \hbox{$V_\mathrm{xs}\!=\!16.7(1)$ $E_\mathrm{r}$}, \hbox{$V_\mathrm{ys}\!=\!17.2(1)$ $E_\mathrm{r}$}, $J_\mathrm{x}/U\!=\!0.12$, $J_\mathrm{y}/U\!=\!0.11$; and \textbf{(d)} \hbox{$V_\mathrm{xs}\!=\!19.2(1)$ $E_\mathrm{r}$}, \hbox{$V_\mathrm{ys}\!=\!18.2(1)$ $E_\mathrm{r}$}, $J_\mathrm{x}/U\!=\!0.064$, $J_\mathrm{y}/U\!=\!0.075$. The measured data for spin imbalance are fitted with a damped sine wave (blue lines).  Shown in the insets are the fast decay processes in the beginning of the dynamics (red circles), matching well the theoretical predictions by including the noises of vacant fillings (red lines). The error bars denote statistical errors, which are $\pm 1\sigma$.}
\end{figure*}

More generally, within the subspace of \hbox{$\mathbb{H}\!:\!\{|\!\uparrow,\downarrow,\uparrow,\downarrow\rangle$}, \hbox{$ |\!\downarrow,\uparrow,\downarrow,\uparrow\rangle\}$}, the Hamiltonian $\hat{H}_{\sss \mathrm{R}}$ (Eq.1) is equivalent to a minimal instance of the Kitaev's toric code model \cite{kitaev2003fault},
\begin{align}\label{eqnkitaev}
\hat{H}_{\sss \mathrm{T}}\!=\!-J_{ \sss \square} \sigma_1^x \sigma_2^x \sigma_3^x \sigma_4^x  -\frac{J_{\sss  +}}{4}\sum_{\langle j,k\rangle} \sigma_j^z\sigma_k^z,
\end{align}
where $\sigma^{x(z)}_j$ are the pauli operators on site $j$.  In the context of  the toric code model, the ground state is a GHZ-type state \hbox{$| \square\rangle \!=\! (|\!\uparrow,\uparrow,\uparrow,\uparrow\rangle\! +\! |\!\downarrow,\downarrow,\downarrow,\downarrow\rangle)/\sqrt{2}$}, while the two antiferromagnetically ordered states $|A^{\sss +}\rangle$ and $|A^{\sss -}\rangle$ are energetically high lying eigenstates. The excited state with an energy of $2J_{ \sss \square}$ takes the form \hbox{$|\boxdot\rangle \!=\!( |\!\uparrow,\uparrow,\uparrow,\uparrow\rangle\!-\!|\!\downarrow,\downarrow,\downarrow,\downarrow\rangle  )/\sqrt{2}$}. It can be generated by applying, for example, the $\sigma^z_1$ operation to the ground state, and is known as a quasiparticle of ``electric charge'' ($e$-particle). Another type of excitation with an energy of $2J_{ +}$ can be generated by applying a single-qubit rotation of $\sigma_j^{x}$ to $| \square\rangle$, called ``magnetic vortices'' (\textit{m}-particles).  These two types quasiparticles are relative $\frac{1}{2}$-anyons of the Toric code model. Their anyonic fractional statistics can be demonstrated by a nontrivial phase factor acquired after a braiding operation, i.e. moving an \textit{m}-particle around an \textit{e}-particle, or vice versa. Note that equations (\ref{eqnring}) and (\ref{eqnkitaev}) for fermions have the similar forms as for bosons except that the signs before \hbox{$J_{\sss +}$} are positive and $|A^{\sss +}\rangle$ and $|A^{\sss -}\rangle$ will be the ground state and the first excited state of the system respectively.

To study the four-body ring-exchange interactions, we initially prepare an ultracold ensemble of $^{87}$Rb atoms with two relevant internal states \hbox{$|\!\downarrow\rangle\!=\!|F=1,m_{ \mathrm{F}}\!=\!-1\rangle$} and \hbox{$|\!\uparrow\rangle\!=\!|F=2,m_{ \mathrm{F}}\!=\!-2\rangle$} in a two-dimensional (2D) array of disconnected plaquettes into an antiferromagnetically ordered configuration \hbox{$|\!\downarrow,\uparrow,\downarrow,\uparrow\rangle$} (Fig.2a). The state initialization is started by preparing a 2D Mott insulator in a square lattices \cite{dai2015generation}. Thereafter, the plaquette potentials are generated by two orthogonal spin-dependent superlattices in the $xy$ plane as shown in Fig.1a. Each of them is created by overlapping two standing waves with periodicity $383.5$ nm (short lattice) and $767$ nm  (long lattice) with a balanced structure. The depths of the short lattices ($V_{\mathrm{xs}}$ and $V_{\mathrm{ys}}$) and long lattices ($V_{\mathrm{xl}}$ and $V_{\mathrm{yl}}$) are given in units of the recoil energy \hbox{$E_{\mathrm{r}}\!=\!h^2/(2m\lambda^2)$}, where $\lambda\!=\!767$ nm, $m$ is the mass of the atom and $h$ is the Plank constant.
With a bias magnetic field of $1.39$ G along \hbox{$\hat{x}\!+\!\hat{y}$} direction, we use a resonant microwave (MW) field \hbox{$\sim\!6.8$ GHz} to couple the spin states \hbox{$|\!\downarrow\rangle$} and \hbox{$|\!\uparrow\rangle$}. The degeneracy of the transitions \hbox{$|\!\downarrow\rangle \!\rightarrow\! |\!\uparrow\rangle$} on each plaquette site is removed by setting the lattice depths \hbox{$V_\mathrm{xs} \!=\! V_\mathrm{ys} \!=\! 150(1)$ $E_\mathrm{r}$}, \hbox{$V_\mathrm{xl} \!=\! V_\mathrm{yl} \!=\! 56.3(4)$} and tuning the short-lattice polarization intersections to $\pi/3$ (see supplementary materials). Thereby a spin-dependent potential is created, \hbox{${\it \Delta}_1\!=\!-32.6(1)$ kHz}, \hbox{${\it \Delta}_3\!=\!32.6(1)$ kHz}, and \hbox{${\it \Delta}_2\!=\!{\it \Delta}_4=0$}.
Hence, we selectively transfer the atoms on sites 2 and 4 into \hbox{$|\!\uparrow\rangle$} via a resonant MW $\pi$-pulse of $60.5$ $\mu$s (Fig.2a). As a result, the spin configuration in every plaquette is initialized to the antiferromagnetically ordered state $|\!\downarrow,\uparrow,\downarrow,\uparrow\rangle$ in parallel. An overall state initialization efficiency of $92(2)\%$ is derived by applying another addressing $\pi$-pulse after $200$ ms holding in the lattice, and measuring the residual atoms on \hbox{$|\!\uparrow\rangle$}. The same selective addressing procedure is used in the spin-imbalance measurement in the later experiment.

\begin{figure}[!t]
\centerline{\includegraphics[width=\linewidth]{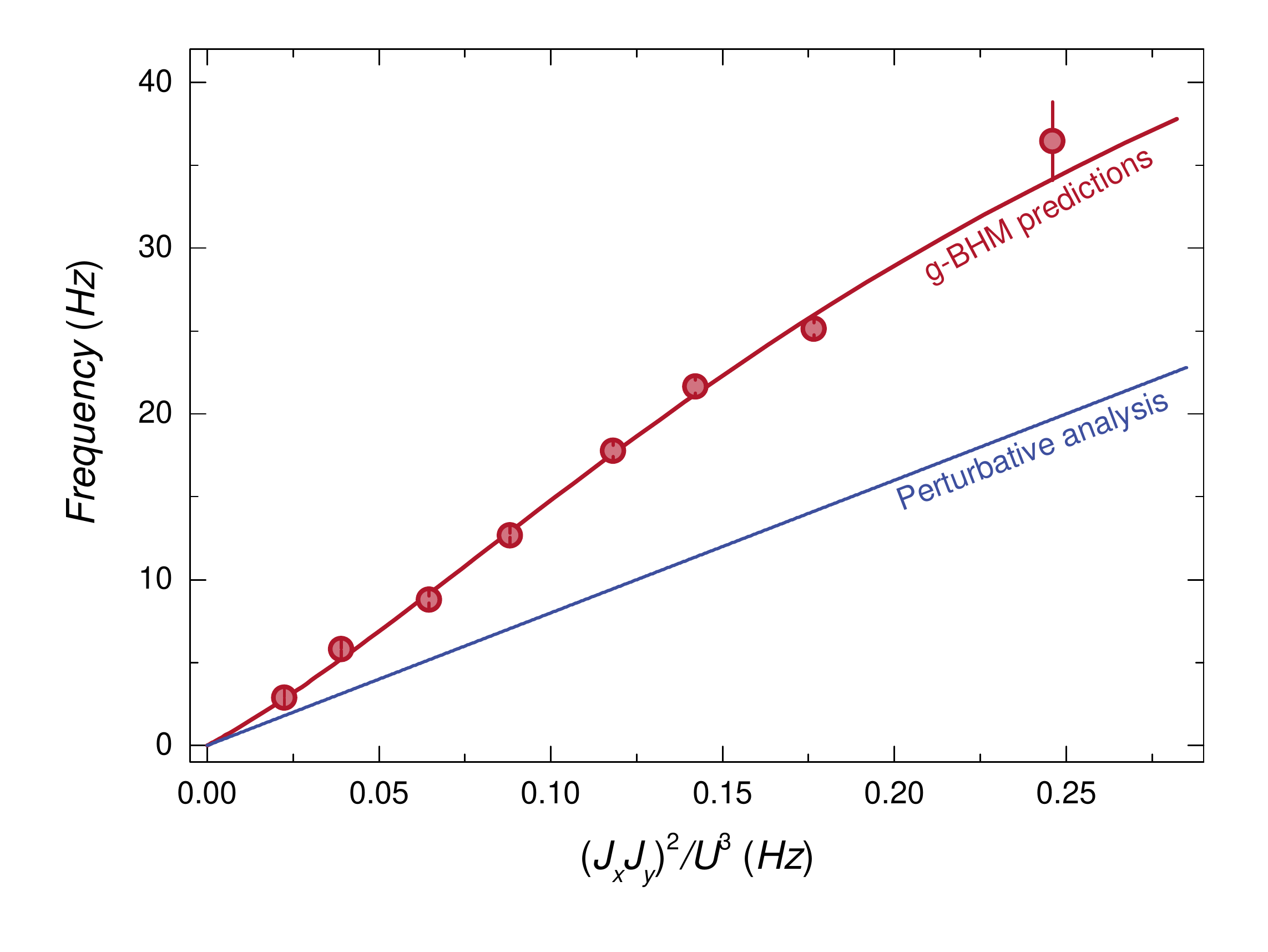}}
\caption{Frequencies of the spin oscillations. The frequencies obtained by fitting the spin imbalanced data for various values of $V_\mathrm{xs}$ and $V_\mathrm{ys}$ (red circles) are compared to the theoretical predictions of the perturbative analysis (blue line) and a generalized BHM (red line). 
}
\end{figure}

The four-spin dynamics can be controlled by the tunneling parameters $J_{\mathrm{x(y)}}$, the onsite interactions $U$ and the intra-plaquette gradients ${\it \Delta}_\mathrm{x(y)}$.
We first ramp down the long lattices to $10$ $E_\mathrm{r}$, and set \hbox{${\it \Delta}_\mathrm{x}\!=\! 115(1)$} Hz, \hbox{${\it \Delta}_\mathrm{y}\!=\! 145(1)$} Hz.
Then the four-spin dynamics in the plaquette are initiated by rapidly ramping down the short lattices within $500$ $\mu$s. After letting the system evolve for a time $t$, we halt the ring-exchange driven dynamics and freeze the spin configurations by simultaneously ramping up the short lattices to $150$ $E_\mathrm{r}$  in $1.5$ ms. We detect the four-spin configurations by applying a site-resolved spin flipping operation, and taking two absorption images $N_\uparrow(\mathrm{x},\mathrm{y})$ and $N_\downarrow(\mathrm{x},\mathrm{y})$ in sequence (Fig.2b). The two plaquette configurations \hbox{$|\!\uparrow,\downarrow,\uparrow,\downarrow\rangle$} and \hbox{$|\!\downarrow,\uparrow,\downarrow,\uparrow\rangle$} are respectively recorded. Therefore the spin imbalance can be derived by
\begin{align}
N_z = \frac{ \sum_{\sss \mathrm{ROI}}\left( N_\uparrow(\mathrm{x},\mathrm{y})- N_\downarrow(\mathrm{x},\mathrm{y})\right)}
{\sum_{\sss \mathrm{ROI}}\left( N_\uparrow(\mathrm{x},\mathrm{y})+N_\downarrow(\mathrm{x},\mathrm{y})\right)},
\end{align}
where a region of interest (ROI) in the center of the atom cloud containing about 272 plaquettes is selected for data analysing.

Two typical evolution curves are shown in Fig.2c-d, where the slow oscillations of $N_z(t)$ reveals the corresponding four-body ring-exchange interactions. A ring-exchange oscillation frequency of $21.6(4)$ Hz is obtained under the Hubbard parameters \hbox{$J_\mathrm{x}/U \!=\! 0.12$}, \hbox{$J_\mathrm{y}/U \!=\! 0.11$}. The emergence of some higher frequency components indicates that the lower order processes have not been fully suppressed.
Increasing the barriers of the plaquette, a lower ring-exchange frequency of $2.9(1)$ Hz is derived under the condition of \hbox{$J_\mathrm{x}/U\!=\!0.064$}, \hbox{$J_\mathrm{y}/U\!=\!0.075$}, where the oscillation becomes smoother and the higher frequency components becomes invisible.
The finite amplitudes \hbox{$0.26(3)$} of the oscillations are mainly limited by the Mott filling properties, which give rise to \hbox{$\sim\!0.37$} of the full amplitude. Further reductions can be explained by the imperfections in the efficiencies of the microwave addressing operations and the suppression of the lower order processes. The background noises caused by the plaquettes with defects decay to a steady value of $N_z$ at the beginning of evolutions through the bare tunneling processes (see insets in Fig.2c, d and supplementary materials).
The damping of the ring-exchange oscillations is induced by the coupling with other magnetically sensitive states which are not in the subspace of $\mathbb{H}$, as well as the dephasing caused by the spatial inhomogeneities of coupling parameters within the ensemble.

The four-body ring-exchange evolutions are measured at varying parameters as shown in Fig.3,  with the oscillation frequencies ranging from 2.9(1) Hz to 36(2) Hz. A comparison between the experiment and the theoretical predictions of the generalized Bose-Hubbard model (g-BHM) \cite{dutta2015non} results in an excellent agreement. To form a g-BHM based on the simple BHM, we introduce modifications including the density-induced tunneling, nearest-neighbor interactions and pair tunneling terms (see supplementary materials). The deviation from the perturbation analysis is due to the imperfect suppression of the lower-order processes. However, the ring-exchange interaction are always the dominant term and the eigenstates are highly projected to the $\{| A^{\sss +}\rangle,|A^{\sss -}\rangle\}$ states.

\begin{figure}[!t]
\centerline{\includegraphics[width=\linewidth]{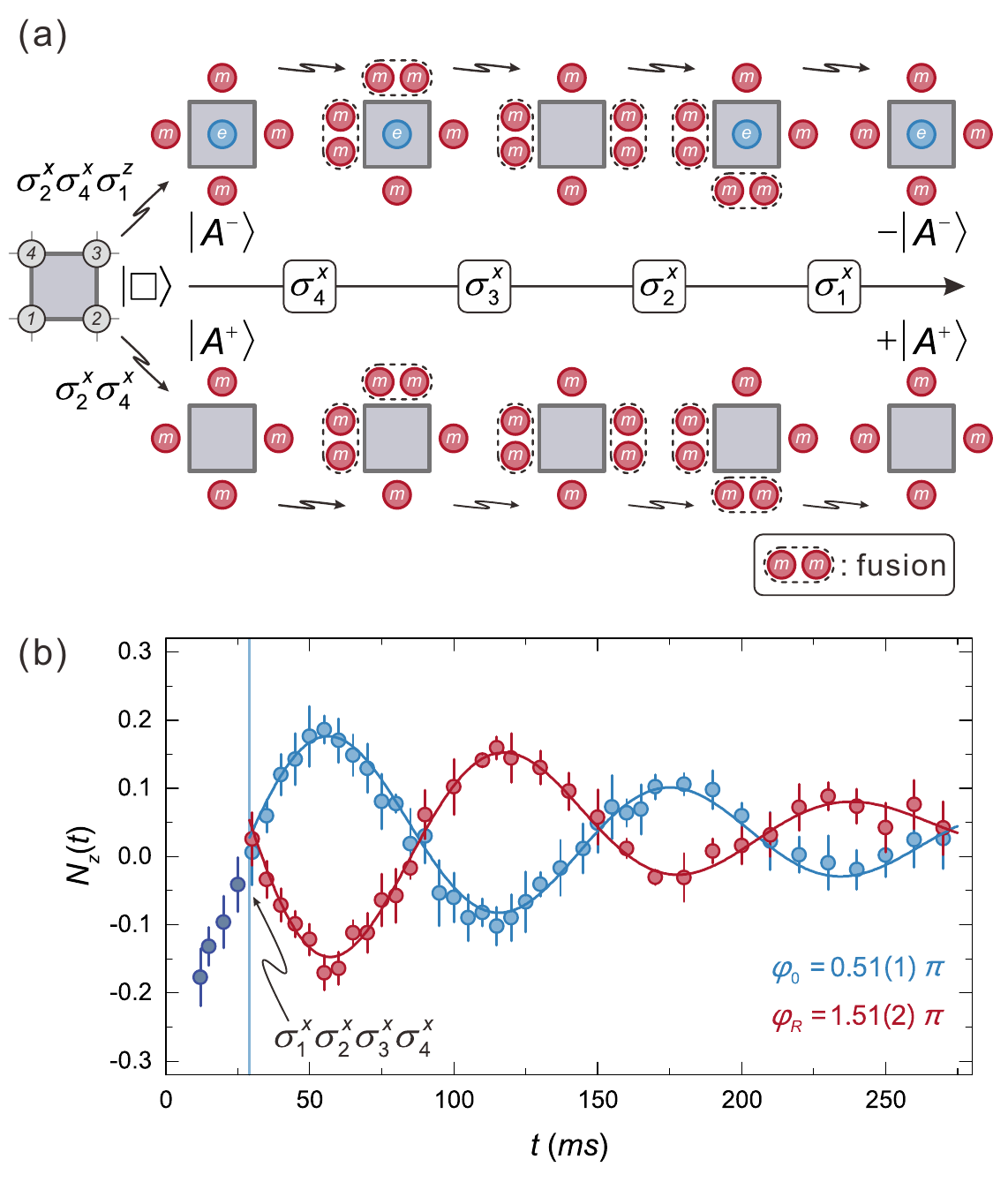}}
\caption{Anyonic fractional statistics. \textbf{(a)} Anyon braiding operations on quasiparticles: first two $m$-particles are created by $\sigma_4^x$, then a single $m$-particle is moved around the $e$-particle by subsequent application of $\sigma_3^x$, $\sigma_2^x$ and $\sigma_1^x$. The two $m$-particles are annihilated in the end, and the system picks up a nontrivial phase factor of \hbox{$-1$}. For the case without the presence of $e$-particle, no additional phase is acquired. \textbf{(b)} Two distinct ring-exchange driven evolutions of the spin imbalance with (red circles) and without (blue circles) implementing the braiding operation at $T/4$ are measured,  with the setting of \hbox{$V_\mathrm{xs}\!=\! 18.2$ $E_\mathrm{r}$}, \hbox{$V_\mathrm{ys}\!=\! 17.2$ $E_\mathrm{r}$}, \hbox{$J_\mathrm{x}/U\!=\!0.082$}, \hbox{$J_\mathrm{y}/U\!=\!0.10$}, \hbox{$T\!=\!117(1)$} ms. Fitting the results with damped sine waves (red and blue lines) and comparing the phase parameters, a relative phase shift of ${\it \Delta}\varphi\!=\!1.00(3)\pi$ is derived after the braiding operation. 
}
\end{figure}

By utilizing the ring-exchange dominating Hamiltonian $\hat{H}_{\sss \mathrm{R}}$ and employing braiding operations, the anyonic quasiparticles of the toric code model (\hbox{Eq.2}) can be artificially created and demonstrated.
The two states \hbox{$|A^{\sss \pm}\rangle$} involved in the ring-exchange dynamics above can be generated by spin operations on the ground states of the toric code model, i.e. \hbox{$|A^{\sss +}\rangle\!=\! \sigma_{2}^x\sigma_{4}^x|\square\rangle$} and \hbox{$|A^{\sss -}\rangle\!=\!\sigma_{2}^x\sigma_{4}^x\sigma_{1}^z|\square\rangle$}. Hence, there are four $m$-particles exist on the four edges of the plaquette in $|A^{\sss +}\rangle$, while an additional $e$-particle lives at the center of the plaquette in $|A^{\sss -}\rangle$. Based on the fusion rules, when two anyons of the same type are created at the same position, they annihilate \cite{pachos2009revealing}. For example, \hbox{$\sigma_1^{x}\sigma_1^{x}|\square\rangle \!=\! |\square\rangle$}.
The braiding on the anyonic quasiparticles can be described as follows (Fig.4a): first creating a pair of $m$-particles on the two edges of the plaquette, then cyclicly moving one $m$-particle around an $e$-particle, and annihilating the two $m$-particles in the end, i.e. \hbox{$\sigma_1^x\sigma_2^x\sigma_3^x\sigma_4^x|A^{\sss -}\rangle \!=\! -|A^{\sss -}\rangle$}.  While without the presence of the $e$-particle, the braiding operations do not give rise a phase to the state, as \hbox{$\sigma_1^x\sigma_2^x\sigma_3^x\sigma_4^x| A^{\sss +}\rangle \!=\! | A^{\sss +}\rangle$}. The nontrivial phase factor of $-1$ acquired after the braiding operations denotes the presence of relative $\frac{1}{2}$-anyons.

To reveal the fractional statistics between the anyonic quasiparticles, we employ an anyonic interferometer \cite{paredes2008minimum} to obtain the phase change by the braiding operations. We first prepare the system to a superposition state \hbox{$\left( | A^{\sss +}\rangle \!+\! i| A^{\sss -} \rangle \right)/\sqrt{2}$} through a quarter period of ring-exchange oscillation (Fig.4b). Then a MW $\pi$-pulse of 19 $\mu$s is applied to flip every site on the plaquette, which is equivalent to perform the braiding operations $\sigma_1^x \sigma_2^x \sigma_3^x \sigma_4^x$ on the four-spin states. The state $|A^{\sss -}\rangle$ picks up an additional phase factor $-1$ because of $\frac{1}{2}$-anyonic statistics, and the system ends up in \hbox{$\left( | A^{\sss +}\rangle\! -\! i|A^{\sss -}\rangle \right)/\sqrt{2}$}. Afterwards, the state evolves under the same ring-exchange setting and the dynamical evolution  is recorded (Fig.4b). A phase shift \hbox{$\Delta\varphi\!=\!\varphi_{\sss \mathrm{R}}\! -\! \varphi_0\!=\!1.00(3) \pi$} can be obtained by comparing the two oscillations with and without implementing the braiding operations. Thus the phase factor \hbox{$e^{i\Delta\varphi}\!=\!-1.00(3)$} acquired on $|A^{\sss -}\rangle$ after braiding is obtained.
Note that the MW $\pi$-pulse is shorter than the ring-exchange period by four orders of magnitude, thus the braiding operations add negligible evolving phases to the state.
Consequently, the observed phase change proves the anyonic fractional statistics in the quasiparticles of the toric code model.

In summary, we have engineered the Hubbard parameters of a plaquette system and achieved a minimum instance of the toric-code Hamiltionian. The energy gap, originating from the ring-exchange interaction, between two anyonic excitations has been observed by measuring dynamical oscillations between the spin configurations \hbox{$|\!\downarrow,\uparrow,\downarrow,\uparrow \rangle$} and \hbox{$|\!\uparrow,\downarrow,\uparrow,\downarrow\rangle$}. Applying a braiding operation to the plaquette state, we found an accumulated phase $\pi$, demonstrating the exotic fractional statistics of Abelian anyonic excitations \cite{kitaev2003fault}. Further cooling the atoms favors the creation of plaquette arrays with less defects and a variety of strongly correlated many-body states  \cite{WenBookQFT,AltmanPRB2002,buchler2005atomic} could be achieved by appropriately connecting the plaquettes  \cite{paredes2008minimum}. Inter-plaquette coupling may be induced by lowering down the barriers or using dynamically driven lattices \cite{MorschRMP2006}. The present method can be adapted to study non-Abelian anyons in honeycomb lattices for topological quantum computation  \cite{kitaev2006anyons,duan2003controlling, ZhangPNAS2007,nayak2008non}. Our experiment represents an essential step towards engineering topological quantum matters with ultracold atoms for quantum simulation and quantum computation.

\bibliographystyle{apsrev4-1} 
\bibliography{reference}

\clearpage
\newpage

\setcounter{equation}{0}
\setcounter{figure}{0}
\setcounter{table}{0}
\setcounter{page}{1}
\makeatletter
\makeatother
\global\def\theequation{S\arabic{equation}}
\global\def\thefigure{S\arabic{figure}}%
\global\def\thetable{S\arabic{table}}
\global\def\thepage{S\arabic{page}}

\onecolumngrid

\begin{center}
{\Large \bf{Supplemental Materials}}
\end{center}

\section*{Experimental Apparatus}

\subsection*{Preparing a two-dimensional Mott insulator state}
The procedure of preparing a two-dimensional Mott insulator is similar as in Ref. \cite{dai2015generation}. Our experiment starts by a $^{87}$Rb Bose-Einstein condensate (BEC) of typical \hbox{$2\!\times\! 10^5$}  atoms in  \hbox{$|\!\downarrow\rangle\!=\!|F\!=\!1,m_\mathrm{F}\!=\!-1\rangle$} without detectable thermal fraction. The atom cloud is then loaded into one layer of a ``pancake" lattice with 4 $\mu$m spatial period, resulting in a two-dimensional quantum gas of \hbox{$1\!\times\!10^4$} atoms with a tight confinement of trapping frequency $6.55(1)$ kHz along the $\hat{z}$ direction. Afterwards we adiabatically ramp up both $\hat{x}$ and $\hat{y}$ short lattices ($\lambda$ = 767 nm) to 25 $E_\mathrm{r}$, the quantum gas undergos a phase transition from a superfluid phase to a Mott insulator. By performing \emph{in-situ} absorption imaging with a spatial resolution of \hbox{$\sim\!2$} $\mu$m, we obtain the atomic density distribution of the ensemble. The central part of the atom cloud containing $272$ plaquettes are chosen as the region of interest (ROI) for data processing.

The filling factor of the Mott insulator is measured using the technique of hyperfine changing collisions \cite{dai2015generation}. We first ramp up the depth of the short lattices to 60 $E_\mathrm{r}$ to inhibit the atom tunneling, then use a $\sim$6.8 GHz microwave (MW) $\pi$-pulse to transfer all the atoms from \hbox{$|\!\downarrow\rangle$} to a hyperfine Zeeman state \hbox{$|F\!=\!2,m_F\!=\!-1\rangle$}. If there are two atoms in the same lattice site, these atoms would experience a spin changing collision process and escape from the lattice confinement. By monitoring the atom loss in short lattices we get the ratio of double occupancies $p_2$. Subsequently, we utilize the same method to the nearest-neighbour lattice sites by transferring  the atoms from the short lattice to a long lattice in one direction, e.g. $\hat{x}$-direction. Therefore obtain the double fillings in the long lattice which are the square of the single fillings in short lattices, $p_1^2$. This two-stage detection reveals that \hbox{$p_1\!=\!73\%$} of the lattice sites in the ROI are filled with one atom, \hbox{$p_2\!\approx\!2\%$} are filled with two atoms and the rest are vacancy defects.


\subsection*{Spin-dependent plaquette lattice and site-selectively addressing}
\begin{figure}[b]
\centerline{\includegraphics[width=9.87cm]{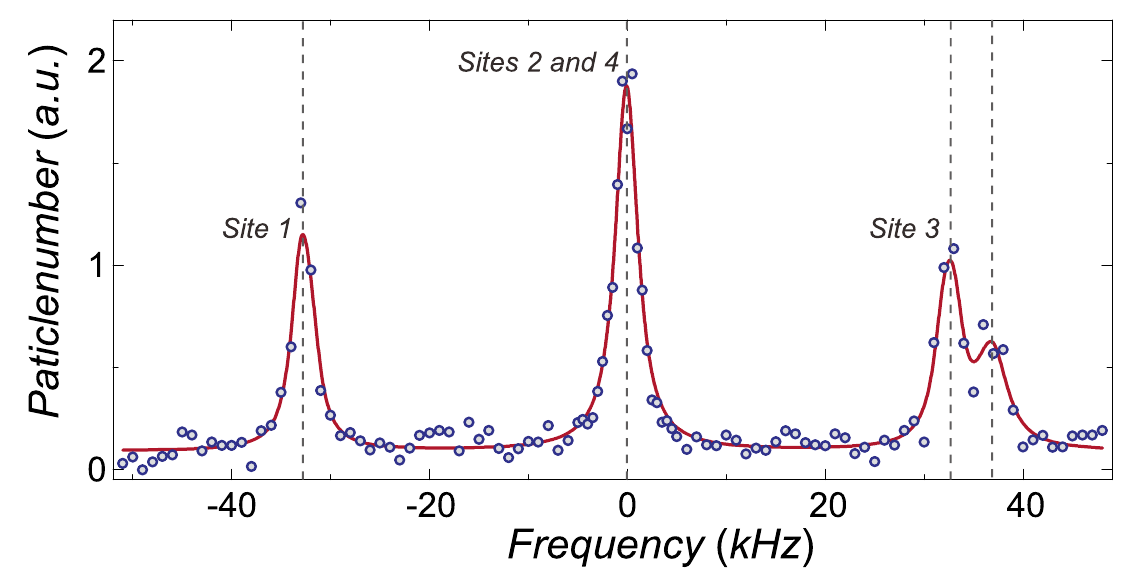}}%
\caption{\label{SFig:spectrum}Site-resolved microwave addressing spectroscopy. By fitting the normalized particle number with a Lorentzian function, the transition frequency of site 1 and 3 are relatively shifted by -32.6(1) kHz and +32.6(1) kHz respectively. The last peak at +36.9(2)kHz corresponds to atoms on site 1 transferred to the first-excited Bloch band.}
\end{figure}

The spin-dependent superlattice is created by modulating the polarization of short lattice lasers \cite{dai2015generation}. Its strength are controlled by the intersection angle $\theta_{x(y)}$ between the incident and retro-reflected laser polarizations through an electro-optic modulator. Hence, it produces a local effective magnetic gradient for the two states \hbox{$|\!\downarrow\rangle$} and \hbox{$|\!\uparrow\rangle\!=\!|F\!=\!2,m_\mathrm{F}\!=\!-2\rangle$}. We set the magnetic quantization axis along the \hbox{$\hat{x}\! -\! \hat{y}$} direction, leading to the intra-plaquette gradients ${\it \Delta}_x$ and ${\it \Delta}_y$ along the $\hat{x}$- and $\hat{y}$-direction, respectively.

To initialize the antiferromagnetically ordered state \hbox{$|\!\downarrow,\uparrow,\downarrow,\uparrow\rangle$} within a plaquette, we implement a strong magnetic gradient and selectively flip the spins on each site by MW pulses. In the condition of lattice depths \hbox{$V_\mathrm{xs}\! = \!V_\mathrm{ys} \!=\!150$} $ E_\mathrm{r}$, \hbox{$V_\mathrm{xl}\! =\!V_\mathrm{yl} \!=\! 56.3$} $ E_\mathrm{r}$ and \hbox{$\theta _{\mathrm{x}}\!=\!  \theta _{\mathrm{y}}\!=\! \pi/3$}, the splitting in both direction is \hbox{${\it \Delta}_\mathrm{x} \!=\! {\it \Delta}_\mathrm{y} \!=\!16.3$} kHz. Thus, the induced magnetic gradient is along the \hbox{$\hat{x} \!-\! \hat{y}$} direction, i.e. sites 2 and 4 have the same resonant MW frequency. We actively lock the magnetic bias field at 1.39 G, and get a spectrum of flipped atoms by scanning the frequency of a weak 500 $\mu$s-length MW pulse (shown in Fig.\ref{SFig:spectrum}). The resonant transition frequencies on sites 1 and  3 are shifted -32.6(1) kHz and +32.6(1) kHz with respect to that of sites 2 and 4. During the state initialization, we use a stronger resonant MW $\pi$-pulse with 8.3 kHz Rabi frequency to flip the spins on sites 2 and 4, resulting in a state preparation efficiency of 92(2)$\%$. The detection of spin configurations also starts from the same site-resolved addressing procedure. Moreover, the intra-plaquette gradients of \hbox{${\it \Delta}_x\! = \! 115(1)$} Hz and \hbox{${\it \Delta}_y \!=\! 145(1)$} Hz are generated during the ring-exchange oscillations to suppress the nearest-neighbor spin-exchange process. They are calibrated by measuring the frequencies of the singlet-triplet oscillations \cite{dai2015generation} between the two Bell states \hbox{$\left(|\!\uparrow,\downarrow\rangle \!+ \!|\!\downarrow,\uparrow\rangle\right)/\sqrt{2}$} and \hbox{$\left(|\!\uparrow,\downarrow\rangle \!-\! |\!\downarrow,\uparrow\rangle\right)/\sqrt{2}$}.

\subsection*{Calibrating the lattices}
To calibrate the lattice depth, we modulate the lattice intensity and then measure the parametric excitations of the atoms from the ground Bloch band to the second excited band. The atoms are loaded into optical lattices with the $\hat{z}$ lattice off to reduce the on-site interaction. We calibrate the depths of the short and long lattices around $40$ $E_\mathrm{r}$ and $22.5$ $E_\mathrm{r}$ respectively. The modulation is kept on for 100 cycles with a depth of  approximately $10\%$ of the laser intensity. We use a Gaussian fit of parametric excitation spectrum to determine the resonant frequency, resulting in a relative accuracy of the peak centers less than $10^{-3}$.
The lattice depths in experiment are then set referring to the calibrated factors. The Hubbard parameters at different lattice depths are derived from numerical calculations of the localized wannier functions in the ground band.


\subsection*{Derive the spin imbalance parameter in experiment}

The characterizing parameter of the ring-exchange dynamics is the spin imbalance in the plaquette, \hbox{$N_z\!=\! \langle \hat{S}^z_1\!-\!\hat{S}^z_2\!+\!\hat{S}^z_3\!-\!\hat{S}^z_4 \rangle/2$}. For the two antiferromagnetically ordered states \hbox{$|\!\uparrow,\downarrow,\uparrow,\downarrow\rangle$} and \hbox{$|\!\downarrow,\uparrow,\downarrow,\uparrow\rangle$}, we will have expectation values of $+1$ and $-1$, respectively. In experiment, the expectation value of $\hat{S}^z_j$ can be derived by the spin occupations $n_{\sigma,j}$, therefore resulting in the spin imbalance
\begin{align}
N_z = & \frac{1}{4} \left( \frac{n_{\uparrow,1}-n_{\downarrow,1}}{n_{\uparrow,1}+n_{\downarrow,1}} - \frac{n_{\uparrow,2}-n_{\downarrow,2}}{n_{\uparrow,2}+n_{\downarrow,2}} + \frac{n_{\uparrow,3}-n_{\downarrow,3}}{n_{\uparrow,3}+n_{\downarrow,3}} - \frac{n_{\uparrow,4}-n_{\downarrow,4}}{n_{\uparrow,4}+n_{\downarrow,4}} \right) \nonumber\\
\approx & \frac{(n_{\uparrow,1}+n_{\downarrow,2} +n_{\uparrow,3}+n_{\downarrow,4})-(n_{\downarrow,1} + n_{\uparrow,2} + n_{\downarrow,3} + n_{\uparrow,4})}{4\bar{n}},
\end{align}
where the atom filling parameter on each site is assumed to be equal,  \hbox{$ n_{\uparrow,j}\!+\!n_{\downarrow,j}\!=\! \bar{n} $}.

In the experiment, we first selectively flip the spin states in sites 2 and 4 after freezing the spin dynamics in the plaquette, therefore transfer the two antiferromagnetically ordered states into \hbox{$|\!\uparrow,\uparrow,\uparrow,\uparrow\rangle$} and \hbox{$|\!\downarrow,\downarrow,\downarrow,\downarrow\rangle$}, respectivley. Then two absorption images $N_\uparrow(\mathrm{x},\mathrm{y})$ and $N_\downarrow(\mathrm{x},\mathrm{y})$  are taken in sequence to record the occupations of \hbox{$|\!\uparrow\rangle$} and \hbox{$|\!\downarrow\rangle$}, during which the deep short lattices are always on. Taking the mean value in the region of interest (ROI) as the expectations of the spin occupations, we get
\begin{align}
(\bar{n}_{\uparrow,1}+\bar{n}_{\downarrow,2} +\bar{n}_{\uparrow,3}+ \bar{n}_{\downarrow,4})  =  \frac{1}{M} \sum_{\sss \mathrm{(x,y)\in ROI}} N_\uparrow(\mathrm{x},\mathrm{y}), \\
(\bar{n}_{\downarrow,1} + \bar{n}_{\uparrow,2} + \bar{n}_{\downarrow,3} + \bar{n}_{\uparrow,4})  =   \frac{1}{M} \sum_{\sss \mathrm{(x,y)\in ROI}} N_\downarrow(\mathrm{x},\mathrm{y}),\\
4\bar{n}  = \frac{1}{M} \sum_{\sss \mathrm{(x,y)\in ROI}} \left( N_\uparrow(\mathrm{x},\mathrm{y}) +  N_\downarrow(\mathrm{x},\mathrm{y}) \right),
\end{align}
where $M$ denotes the number of plaquette in the ROI. Thus the averaged spin imbalance in the ROI can be experimentally derived as
\begin{align}
N_z \bigg |_\mathrm{(x,y)\in ROI} = \frac{\sum_\mathrm{\sss ROI} \left(N_\uparrow(\mathrm{x},\mathrm{y}) -  N_\downarrow(\mathrm{x},\mathrm{y}) \right)}{\sum_\mathrm{\sss ROI} \left(N_\uparrow(\mathrm{x},\mathrm{y}) +  N_\downarrow(\mathrm{x},\mathrm{y}) \right)}.
\end{align}

\section*{Theoretical Model}
\subsection*{The effective Hamiltonian}

The system of ultracold atomic spins in a four-site plaquette can be described by a two-species single-band Bose-Hubbard Hamiltonian
\begin{align}
\hat{H}^{^\mathrm{BHM}}=& -\sum_{\langle j,k\rangle,\sigma}J_{jk,\sigma}\left(\hat{a}^\dagger_{j\sigma} \hat{a}_{k\sigma} + \mathrm{H.c.} \right)  + \frac{1}{2}\sum_{j,\sigma}U_{j,\sigma\sigma}\hat{n}_{j\sigma}(\hat{n}_{j\sigma}-1) + \sum_{j}U_{j,\uparrow\downarrow}\hat{n}_{j\uparrow} \hat{n}_{j\downarrow}\nonumber\\& + \frac{1}{2}\sum_{j}{\it \Delta}_{j} (\hat{n}_{j\uparrow}-\hat{n}_{j\downarrow}) + \sum_{j}\mu_{j} (\hat{n}_{j\uparrow}+\hat{n}_{j\downarrow}),
\end{align}
where $j,k \!\in\! \{1,2,3,4\}$ denote the site of the plaquette, $\sigma \!\in\! \{\uparrow,\downarrow\}$ denotes the spin components, $\hat{a}^\dagger$ and $\hat{a}$ are the bosonic creation and annihilation operators, $\hat{n}$ is the particle number operator, $J_{jk}$ is the matrix element of the nearest-neighbor tunneling, $U$ describes the onsite interaction between two particles, ${\it \Delta}_j$ is the spin-dependent potential biases induced by magnetic gradients $\it \Delta_\mathrm{x(y)}$, and  $\mu_j$ represents the bias energy offsets. In our experiment, the lattice depths for both spin states are almost the same as the lattice lasers are all far detuned, therefore \hbox{$J_\uparrow \!=\! J_\downarrow\!=\!J$}, \hbox{$U_{\uparrow\uparrow} \!=\! U_{\downarrow\downarrow} \!=\! U_{\uparrow\downarrow}\!=\!U$}, \hbox{$U\!=\!g\int\!\mathrm{d}\mathbf{x}\  |w_{j}(\mathbf{x})|^4$}, \hbox{$g\!=\!h^2a_\mathrm{s}/\pi m$} is the effective interaction strength, $w_{j}(\mathbf{x})$ are the localized Wannier functions.

In the low temperature regime, the underlying spin-spin interactions become the dominating dynamics of the Hubbard model. In the limit of Mott insulator with strong interactions \hbox{$J \!\ll\! U$}, our system of a four-site plaquette can be described by the following basis states with the spin occupations \hbox{$\mathcal{P}\! =\!  \{|\sigma_1,\sigma_2,\sigma_3,\sigma_4\rangle\}$}, where \hbox{$\sigma_j =\ \uparrow \mathrm{ or } \downarrow$} with the subscripts denote the sites of the plaquette, and the commas separate the occupations of the four sites.  This basis contains \hbox{$2^4\!=\!16$} states with different spin configurations. Outside of this basis are the states with multiple spins in one site, which have higher energies \hbox{$\propto\!U$}.
\begin{figure}[t]
\centerline{\includegraphics[width=14.5cm]{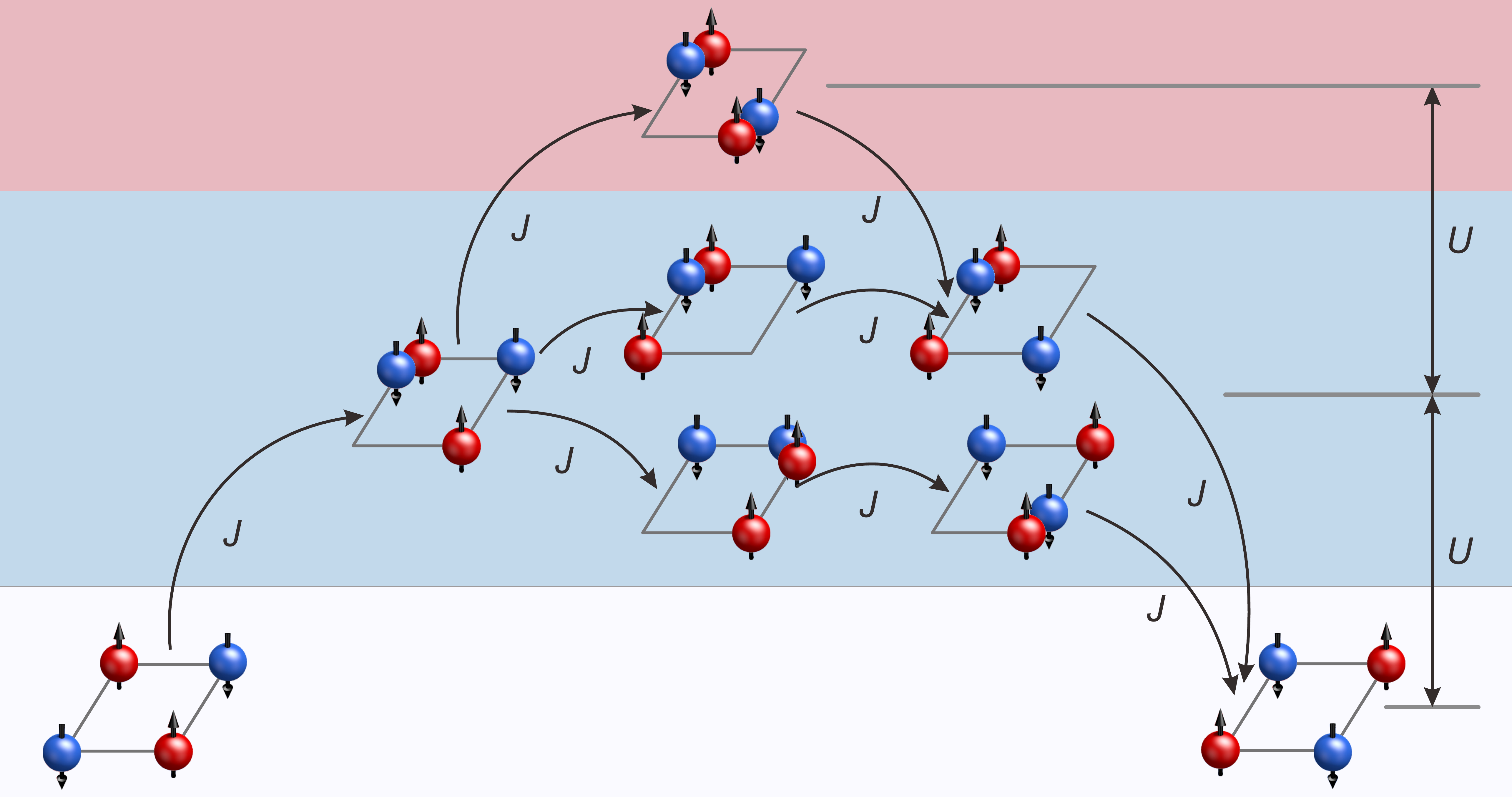}}%
\caption{\label{SFig:perutrbations}Three typical pathways of the ring-exchange processes in the forth order perturbation theory.}
\end{figure}

We evaluate the action of the tunneling operator via the perturbation theory in the strong interaction regime. The first order processes will lead the system out of the basis and therefore are energetically forbidden.  By introducing spin-dependent intra-plaquette gradients with \hbox{${\it \Delta}_\mathrm{x(y)}\!\gg\!4J_\mathrm{x(y)}^2/U$}, the second order processes mediated by the superexchange interactions between neighbouring spins are also suppressed. Although the Ising-type interactions \hbox{$-J_{\sss +}\hat{S^z_j}\hat{S^z_k}$} between the neighbouring spins still exist in the system, they do not couple different spin states. However, the forth order tunneling processes that leave the system within $\mathcal{P}$ lead to an effective coupling between the different spin states. We can describe such processes via an effective Hamiltonian of the system, whose matrix elements within $\mathcal{P}$ can be evaluated via the forth order perturbation theory:
\begin{align}
H_{m,n}^\mathrm{eff} = -\sum_{k_j \not\in \mathcal{P}}\langle m | \hat{H}_J| k_4 \rangle \frac{1}{\langle k_4 | \hat{H}_\mathrm{int} | k_4\rangle} \langle k_4 | \hat{H}_J | k_3\rangle \frac{1}{\langle k_3 | \hat{H}_\mathrm{int} | k_3\rangle} \langle k_3 | \hat{H}_J| k_2 \rangle \frac{1}{\langle k_2 | \hat{H}_\mathrm{int} | k_2\rangle} \langle k_2 | \hat{H}_J | n\rangle.
\end{align}
Here $\hat{H}_J$ denotes the tunnelling and $\hat{H}_\mathrm{int}$ the interaction part of the Hubbard Hamiltonian.

The intra-plaquette gradients remove part of the energy degeneracies between different spin states within $\mathcal{P}$. However, the two antiferromagnetically ordered states \hbox{$|\!\downarrow,\uparrow,\downarrow,\uparrow\rangle$} and \hbox{$|\!\uparrow,\downarrow,\uparrow,\downarrow\rangle$}, which fulfill both spin and energy conservations, can be coupled by a forth order process mediated by the four-body ring-exchange interactions. In the condition of \hbox{$\Delta_\mathrm{x(y)}\! \ll \! U$}, the strength of the ring-exchange interactions $-J_{\sss \square}\!\approx\!-40J_\mathrm{x}^2 J_\mathrm{y}^2 /U^3$
are obtained by summing up all the pathways (Fig.\ref{SFig:perutrbations}). Evaluating the effective matrix elements for this process and diagonalizing the Hamiltonian matrix, one obtain the eigenstates and eigenenergies
\begin{align}
|A^{\sss -}\rangle =& \frac{1}{\sqrt{2}}\left( |\!\uparrow,\downarrow,\uparrow,\downarrow\rangle\! -\! |\!\downarrow,\uparrow,\downarrow,\uparrow\rangle \right), \textrm{ with } E^{\sss -}= + J_{\sss \square},\\
|A^{\sss +}\rangle =&  \frac{1}{\sqrt{2}}\left( |\!\uparrow,\downarrow,\uparrow,\downarrow\rangle\! +\! |\!\downarrow,\uparrow,\downarrow,\uparrow\rangle \right), \textrm{ with }  E^{\sss +}= - J_{\sss \square}.
\end{align}
Thus, the effective ring-exchange interaction can be written as
\begin{align}
\hat{H}^\mathrm{eff}_\mathrm{\sss \square}  = & - J_{\sss \square} (|A^{\sss +}\rangle\langle A^{\sss +}|  - |A^{\sss -}\rangle\langle A^{\sss -}|) \nonumber\\
 = &-J_{\sss \square} (\hat{S}_1^+ \hat{S}_2^- \hat{S}_3^+ \hat{S}_4^- +\hat{S}_1^- \hat{S}_2^+ \hat{S}_3^- \hat{S}_4^+ ),
\end{align}
where $\hat{S}^\pm_j$ are the spin ladder operators in site $j$. Additionally, when the intra-plaquette gradients are along the \hbox{$\hat{x}\!-\!\hat{y}$} direction, i.e. \hbox{${\it \Delta}_\mathrm{x} \!=\! {\it \Delta}_\mathrm{y} $}, a forth order tunneling process mediated by the next-neighboring exchange interactions between sites 2 and 4 can happen, resulting in an effective diagonal exchange interactions \hbox{$\hat{H}^\mathrm{eff}_\mathrm{\sss \times}\!  = \! - J_{\sss \times}(\hat{S}_2^+ \hat{S}_4^- + \hat{S}_2^- \hat{S}_4^+) $}, where \hbox{$J_{\sss \times}\!\propto\! J_\mathrm{x}^2J_\mathrm{y}^2/U^3$} are the corresponding diagonal exchange matrix elements. However, this term can be suppressed by tune the gradients away from the \hbox{$\hat{x}\!-\!\hat{y}$} direction, i.e. \hbox{${\it\Delta}_\mathrm{xy} \!=\! | {\it\Delta}_\mathrm{x} \!-\! {\it\Delta}_\mathrm{y}| \!\gg\!J_\mathrm{x}^2J_\mathrm{y}^2/U^3$}. Including all these interactions discussed here, we obtain the effective ring-exchange dominating Hamiltonian in the plaquette as
\begin{align}
\hat{H}_{\sss \mathrm{R}} = & \ \hat{H}^\mathrm{eff}_\mathrm{\sss  \square} 
+ \hat{H}_\mathrm{\sss Ising} + \hat{V}_\mathrm{\sss Grad}\nonumber\\
=& -J_{\sss \square} (\hat{S}_1^+ \hat{S}_2^- \hat{S}_3^+ \hat{S}_4^- + \mathrm{H.c} )
- \sum_{\langle j,k\rangle}J_{\sss +}\hat{S^z_j}\hat{S^z_k} + \sum_j {\it \Delta_j} \hat{S}^z_j,
\end{align}
where \hbox{$J_+\!=\!4J_\mathrm{x(y)}^2/U$}, \hbox{${\it \Delta}_1\!=\!-{\it \Delta}_\mathrm{x}\!-\!{\it \Delta}_\mathrm{y}$}, \hbox{${\it \Delta}_2\!=\!{\it \Delta}_\mathrm{x}\!-\!{\it \Delta}_\mathrm{y}$}, \hbox{${\it \Delta}_3\!=\!{\it \Delta}_\mathrm{x}\!+\!{\it \Delta}_\mathrm{y}$}, and \hbox{${\it \Delta}_4\!=\!-{\it \Delta}_\mathrm{x}\!+\!{\it \Delta}_\mathrm{y}$}.

\subsection*{Comparison between $\hat{H}_\mathrm{\sss R}$ and the toric code model}

\begin{figure}
\centerline{\includegraphics[width= 9.cm]{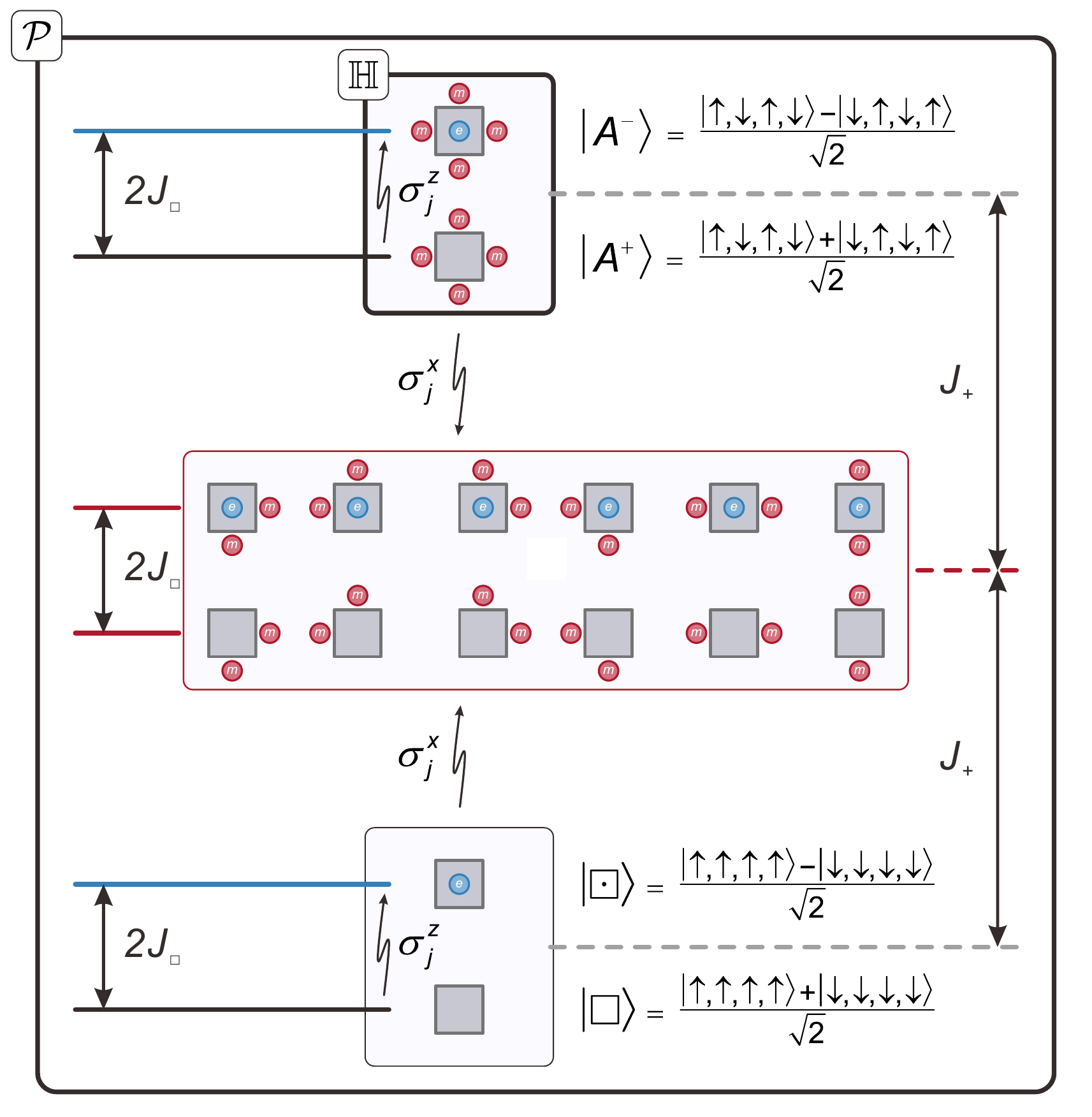}}
\caption{\label{SFig:toric}Eigenstates and eigenenergies in the toric code model $\hat{H}_\mathrm{\sss T}$ within $\mathcal{P}$.}
\end{figure}

\begin{figure}
\centerline{\includegraphics[width= 9.cm]{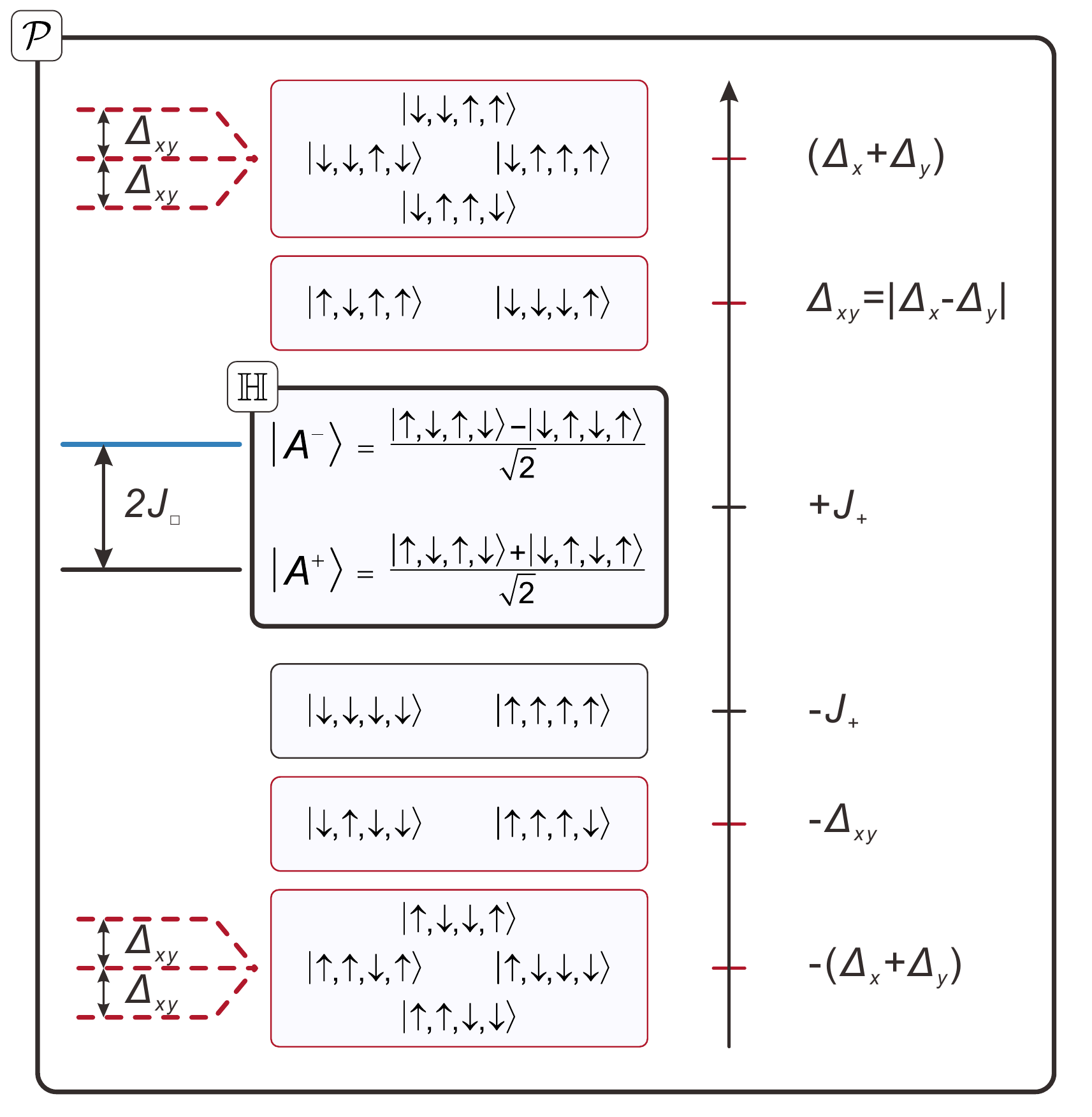}}
\caption{\label{SFig:plaquettemodel}Eigenstates and eigenenergies of the Hamiltonian $\hat{H}_\mathrm{\sss R}$ within $\mathcal{P}$.}
\end{figure}

One can find great similarity between the first two terms of the Hamiltonian $\hat{H}_\mathrm{\sss R}$ between a minimal instance of the Kitaev's toric code model
\begin{align}
\hat{H}_\mathrm{\sss T}\!=\! -\hat{A}_s  - \hat{B}_p,
\end{align}
with \hbox{$\hat{A}_s\!=\! J_{ \sss \square}\sigma_1^x \sigma_2^x \sigma_3^x \sigma_4^x$}, \hbox{$\hat{B}_p\!=\!J_{\sss  +}\sum_{\langle j,k\rangle} \sigma_j^z\sigma_k^z/4$}. We evaluate the eigenstates and eigenenergies of the toric code model in the basis of $\mathcal{P}$ with the bosonic coupling terms \hbox{$-J_{\sss +}\!<\!0$} and \hbox{$-J_{\sss \square}\!<\!0$}, as shown in  Fig.\ref{SFig:toric}. The ground state takes the form \hbox{$|\square\rangle\!=\! (|\!\uparrow,\uparrow,\uparrow,\uparrow\rangle \!+\! |\!\downarrow,\downarrow,\downarrow,\downarrow\rangle)/\sqrt{2}$}, while the two states $|A^{\sss \pm}\rangle$ in the subspace \hbox{$\mathbb{H}\!=\! \{ |\!\downarrow,\uparrow,\downarrow,\uparrow\rangle$}, $|\!\uparrow,\downarrow,\uparrow,\downarrow\rangle\}$ are energetically high lying eigenstates.
Furthermore, we evaluate the eigenstates and eigenenergies of $\hat{H}_\mathrm{\sss R}$ within the basis of $\mathcal{P}$ (Fig.\ref{SFig:plaquettemodel}). The spin-dependent intra-plaquette gradients remove the degeneracies between states except the ones within the subspace $\mathbb{H}$, where the two eigenstates remain in $|A^{\sss \pm}\rangle$ with an energy gap of $2J_{\sss \square}$. Therefore, the two Hamiltonians $\hat{H}_\mathrm{\sss R}$ and $\hat{H}_\mathrm{\sss T}$ are equivalent in the subspace of $\mathbb {H}$.

\subsection*{The generalized Bose-Hubbard model}
Including additional terms of density-induced tunnelings, nearest-neighbor interactions, and the pair tunnelings to the standard BHM Hamiltonian leads to the generalized four-site Bose-Hubbard Hamiltonian (g-BHM)\cite{dutta2015non},
\begin{align}
\hat{H}^{^\mathrm{gBHM}} = & \hat{H}^{^\mathrm{BHM}} -\sum_{\langle j,k \rangle,\sigma}T_{j,k}(\hat{n}_{j\sigma}+\hat{n}_{k\sigma}-1)(\hat{a}_{j\sigma}^\dagger\hat{a}_{k\sigma} + \mathrm{H.c.}) - \sum_{\langle j,k \rangle,\sigma\neq\sigma^{\prime}}T_{j,k}(\hat{n}_{j\sigma}\hat{a}^\dagger_{j\sigma^\prime}\hat{a}_{k\sigma^\prime}+ \hat{n}_{k\sigma}\hat{a}^\dagger_{k\sigma^\prime}\hat{a}_{j\sigma^\prime}+\mathrm{H.c.})\nonumber\\& +\sum_{\langle j,k \rangle,\sigma}2V_{j,k}\hat{n}_{j\sigma}\hat{n}_{k\sigma} +  \sum_{\langle j,k \rangle,\sigma\neq\sigma^{\prime}}V_{j,k}(\hat{n}_{j\sigma}\hat{n}_{k\sigma^{\prime}} + \hat{a}^\dagger_{j\sigma}\hat{a}^\dagger_{j\sigma^{\prime}}\hat{a}_{k\sigma}\hat{a}_{k\sigma^{\prime}} +\mathrm{H.c.})\nonumber\\&
+\sum_{\langle j,k\rangle,\sigma}P_{j,k}\left[(\hat{a}^\dagger_{j\sigma}\hat{a}_{k\sigma})^2+\mathrm{H.c.}\right] +\sum_{\langle j,k\rangle,\sigma\neq\sigma^{\prime}} P_{j,k}( \hat{a}^\dagger_{j\sigma}\hat{a}^\dagger_{j\sigma^\prime}\hat{a}_{k\sigma}\hat{a}_{k\sigma^\prime}+\mathrm{H.c.}),
\end{align}
where \hbox{$T_{\langle j,k\rangle} \!=\! -g\int\!\mathrm{d}\mathbf{x}\  |w_{j}(\mathbf{x})|^2 w^{\ast}_{j}(\mathbf{x})w_{k}(\mathbf{x})$}, \hbox{$V_{\langle j,k\rangle} \!=\! g\int\!\mathrm{d}\mathbf{x}\  |w_{j}(\mathbf{x})|^2|w_{k}(\mathbf{x})|^2$}, \hbox{$P_{\langle j,k\rangle} \!=\! g\int\!\mathrm{d}\mathbf{x}\  w_{j}^\ast(\mathbf{x})^2 w_{k}(\mathbf{x})^2$} are the coupling strengths of density-induced tunnelings, nearest-neighbor interactions, and the pair tunnelings, respectively. In our experiment, the major correction comes from the density-induced tunnelings, which results in an enhancement of \hbox{$3\%\!\sim\!4\%$} to the bare tunneling strength. Therefore, the ring-exchange interactions (\hbox{$\propto\! J^4/U^3$}) estimated by the g-BHM are \hbox{$10\%\!\sim\!15\%$} higher than the predictions by the standard BHM.

\begin{table}[b]
\centering
\begin{tabular}{|c|ccccc|}
\hline
$j$ &   &   & $|\psi_j\rangle\!\in\!\mathcal{H}$ &   &   \\
\hline
1-5 & $|\!\downarrow_{ 1}, \downarrow_{ 1}, \uparrow_{ 1}, \uparrow_{ 1}\rangle$ &$|\!\downarrow_{ 1}, \downarrow_{ 1}, \uparrow_{ 1}, \uparrow_{ 2} \rangle$ &
$|\!\downarrow_{ 1}, \downarrow_{ 1}, \uparrow_{ 1}, \uparrow_{ 3} \rangle$ &$|\!\downarrow_{ 1}, \downarrow_{ 1}, \uparrow_{ 1}, \uparrow_{ 4} \rangle$ &
$|\!\downarrow_{ 1}, \downarrow_{ 1}, \uparrow_{ 2} , \uparrow_{ 2} \rangle$ \\
6-10 & $|\!\downarrow_{ 1}, \downarrow_{ 1}, \uparrow_{ 2} , \uparrow_{ 3} \rangle$ &
$|\!\downarrow_{ 1}, \downarrow_{ 1}, \uparrow_{ 2} , \uparrow_{ 4} \rangle$ &$|\!\downarrow_{ 1}, \downarrow_{ 1}, \uparrow_{ 3} , \uparrow_{ 3} \rangle$ &
$|\!\downarrow_{ 1}, \downarrow_{ 1}, \uparrow_{ 3} , \uparrow_{ 4} \rangle$ &$|\!\downarrow_{ 1}, \downarrow_{ 1}, \uparrow_{ 4} , \uparrow_{ 4} \rangle$ \\
11-15 & $|\!\downarrow_{ 1}, \uparrow_{ 1}, \uparrow_{ 1}, \downarrow_{ 2} \rangle$ & $|\!\downarrow_{ 1}, \uparrow_{ 1}, \downarrow_{ 2} , \uparrow_{ 2} \rangle$ &
$|\!\downarrow_{ 1}, \uparrow_{ 1}, \downarrow_{ 2} , \uparrow_{ 3} \rangle$ &$|\!\downarrow_{ 1}, \uparrow_{ 1}, \downarrow_{ 2} , \uparrow_{ 4} \rangle$ &
$|\!\downarrow_{ 1}, \downarrow_{ 2} , \uparrow_{ 2} , \uparrow_{ 2} \rangle$ \\
16-20 & $|\!\downarrow_{ 1}, \downarrow_{ 2} , \uparrow_{ 2} , \uparrow_{ 3} \rangle$ &
$|\!\downarrow_{ 1}, \downarrow_{ 2} , \uparrow_{ 2} , \uparrow_{ 4} \rangle$ &$|\!\downarrow_{ 1}, \downarrow_{ 2} , \uparrow_{ 3} , \uparrow_{ 3} \rangle$ &
$|\!\downarrow_{ 1}, \downarrow_{ 2} , \uparrow_{ 3} , \uparrow_{ 4} \rangle$ &$|\!\downarrow_{ 1}, \downarrow_{ 2} , \uparrow_{ 4} , \uparrow_{ 4} \rangle$ \\
21-25 & $|\!\downarrow_{ 1}, \uparrow_{ 1}, \uparrow_{ 1}, \downarrow_{ 3} \rangle$ &  $|\!\downarrow_{ 1}, \uparrow_{ 1}, \uparrow_{ 2} , \downarrow_{ 3} \rangle$ &
$|\!\downarrow_{ 1}, \uparrow_{ 1}, \downarrow_{ 3} , \uparrow_{ 3} \rangle$ &  $|\!\downarrow_{ 1}, \uparrow_{ 1}, \downarrow_{ 3} , \uparrow_{ 4} \rangle$ &
$|\!\downarrow_{ 1}, \uparrow_{ 2} , \uparrow_{ 2} , \downarrow_{ 3} \rangle$ \\
26-30 & $|\!\downarrow_{ 1}, \uparrow_{ 2} , \downarrow_{ 3} , \uparrow_{ 3} \rangle$ &
$|\!\downarrow_{ 1}, \uparrow_{ 2} , \downarrow_{ 3} , \uparrow_{ 4} \rangle$ &  $|\!\downarrow_{ 1}, \downarrow_{ 3} , \uparrow_{ 3} , \uparrow_{ 3} \rangle$ &
$|\!\downarrow_{ 1}, \downarrow_{ 3} , \uparrow_{ 3} , \uparrow_{ 4} \rangle$ &  $|\!\downarrow_{ 1}, \downarrow_{ 3} , \uparrow_{ 4} , \uparrow_{ 4} \rangle$ \\
31-35& $|\!\downarrow_{ 1}, \uparrow_{ 1}, \uparrow_{ 1}, \downarrow_{ 4} \rangle$ &  $|\!\downarrow_{ 1}, \uparrow_{ 1}, \uparrow_{ 2} , \downarrow_{ 4} \rangle$ &
$|\!\downarrow_{ 1}, \uparrow_{ 1}, \uparrow_{ 3} , \downarrow_{ 4} \rangle$ &  $|\!\downarrow_{ 1}, \uparrow_{ 1}, \downarrow_{ 4} , \uparrow_{ 4} \rangle$ &
$|\!\downarrow_{ 1}, \uparrow_{ 2} , \uparrow_{ 2} , \downarrow_{ 4} \rangle$ \\
36-40 & $|\!\downarrow_{ 1}, \uparrow_{ 2} , \uparrow_{ 3} , \downarrow_{ 4} \rangle$ &
$|\!\downarrow_{ 1}, \uparrow_{ 2} , \downarrow_{ 4} , \uparrow_{ 4} \rangle$ &  $|\!\downarrow_{ 1}, \uparrow_{ 3} , \uparrow_{ 3} , \downarrow_{ 4} \rangle$ &
$|\!\downarrow_{ 1}, \uparrow_{ 3} , \downarrow_{ 4} , \uparrow_{ 4} \rangle$ &  $|\!\downarrow_{ 1}, \downarrow_{ 4} , \uparrow_{ 4} , \uparrow_{ 4} \rangle$ \\
41-45 & $|\!\uparrow_{ 1}, \uparrow_{ 1}, \downarrow_{ 2} , \downarrow_{ 2} \rangle$ &  $|\!\uparrow_{ 1}, \downarrow_{ 2} , \downarrow_{ 2} , \uparrow_{ 2} \rangle$ &
$|\!\uparrow_{ 1}, \downarrow_{ 2} , \downarrow_{ 2} , \uparrow_{ 3} \rangle$ &  $|\!\uparrow_{ 1}, \downarrow_{ 2} , \downarrow_{ 2} , \uparrow_{ 4} \rangle$ &
$|\!\downarrow_{ 2} , \downarrow_{ 2} , \uparrow_{ 2} , \uparrow_{ 2} \rangle$ \\
46-50 & $|\!\downarrow_{ 2} , \downarrow_{ 2} , \uparrow_{ 2} , \uparrow_{ 3} \rangle$ &
$|\!\downarrow_{ 2} , \downarrow_{ 2} , \uparrow_{ 2} , \uparrow_{ 4} \rangle$ &  $|\!\downarrow_{ 2} , \downarrow_{ 2} , \uparrow_{ 3} , \uparrow_{ 3} \rangle$ &
$|\!\downarrow_{ 2} , \downarrow_{ 2} , \uparrow_{ 3} , \uparrow_{ 4} \rangle$ &  $|\!\downarrow_{ 2} , \downarrow_{ 2} , \uparrow_{ 4} , \uparrow_{ 4} \rangle$ \\
51-55 & $|\!\uparrow_{ 1}, \uparrow_{ 1}, \downarrow_{ 2} , \downarrow_{ 3} \rangle$ &  $|\!\uparrow_{ 1}, \downarrow_{ 2} , \uparrow_{ 2} , \downarrow_{ 3} \rangle$ &
$|\!\uparrow_{ 1}, \downarrow_{ 2} , \downarrow_{ 3} , \uparrow_{ 3} \rangle$ &  $|\!\uparrow_{ 1}, \downarrow_{ 2} , \downarrow_{ 3} , \uparrow_{ 4} \rangle$ &
$|\!\downarrow_{ 2} , \uparrow_{ 2} , \uparrow_{ 2} , \downarrow_{ 3} \rangle$ \\
56-60 & $|\!\downarrow_{ 2} , \uparrow_{ 2} , \downarrow_{ 3} , \uparrow_{ 3} \rangle$ &
$|\!\downarrow_{ 2} , \uparrow_{ 2} , \downarrow_{ 3} , \uparrow_{ 4} \rangle$ &  $|\!\downarrow_{ 2} , \downarrow_{ 3} , \uparrow_{ 3} , \uparrow_{ 3} \rangle$ &
$|\!\downarrow_{ 2} , \downarrow_{ 3} , \uparrow_{ 3} , \uparrow_{ 4} \rangle$ &  $|\!\downarrow_{ 2} , \downarrow_{ 3} , \uparrow_{ 4} , \uparrow_{ 4} \rangle$ \\
61-65 & $|\!\uparrow_{ 1}, \uparrow_{ 1}, \downarrow_{ 2} , \downarrow_{ 4} \rangle$ &  $|\!\uparrow_{ 1}, \downarrow_{ 2} , \uparrow_{ 2} , \downarrow_{ 4} \rangle$ &
$|\!\uparrow_{ 1}, \downarrow_{ 2} , \uparrow_{ 3} , \downarrow_{ 4} \rangle$ &  $|\!\uparrow_{ 1}, \downarrow_{ 2} , \downarrow_{ 4} , \uparrow_{ 4} \rangle$ &
$|\!\downarrow_{ 2} , \uparrow_{ 2} , \uparrow_{ 2} , \downarrow_{ 4} \rangle$ \\
66-70 & $|\!\downarrow_{ 2} , \uparrow_{ 2} , \uparrow_{ 3} , \downarrow_{ 4} \rangle$ &
$|\!\downarrow_{ 2} , \uparrow_{ 2} , \downarrow_{ 4} , \uparrow_{ 4} \rangle$ &  $|\!\downarrow_{ 2} , \uparrow_{ 3} , \uparrow_{ 3} , \downarrow_{ 4} \rangle$ &
$|\!\downarrow_{ 2} , \uparrow_{ 3} , \downarrow_{ 4} , \uparrow_{ 4} \rangle$ &  $|\!\downarrow_{ 2} , \downarrow_{ 4} , \uparrow_{ 4} , \uparrow_{ 4} \rangle$ \\
71-75 & $|\!\uparrow_{ 1}, \uparrow_{ 1}, \downarrow_{ 3} , \downarrow_{ 3} \rangle$ &  $|\!\uparrow_{ 1}, \uparrow_{ 2} , \downarrow_{ 3} , \downarrow_{ 3} \rangle$ &
$|\!\uparrow_{ 1}, \downarrow_{ 3} , \downarrow_{ 3} , \uparrow_{ 3} \rangle$ &  $|\!\uparrow_{ 1}, \downarrow_{ 3} , \downarrow_{ 3} , \uparrow_{ 4} \rangle$ &
$|\!\uparrow_{ 2} , \uparrow_{ 2} , \downarrow_{ 3} , \downarrow_{ 3} \rangle$ \\
76-80 & $|\!\uparrow_{ 2} , \downarrow_{ 3} , \downarrow_{ 3} , \uparrow_{ 3} \rangle$ &
$|\!\uparrow_{ 2} , \downarrow_{ 3} , \downarrow_{ 3} , \uparrow_{ 4} \rangle$ &  $|\!\downarrow_{ 3} , \downarrow_{ 3} , \uparrow_{ 3} , \uparrow_{ 3} \rangle$ &
$|\!\downarrow_{ 3} , \downarrow_{ 3} , \uparrow_{ 3} , \uparrow_{ 4} \rangle$ &  $|\!\downarrow_{ 3} , \downarrow_{ 3} , \uparrow_{ 4} , \uparrow_{ 4} \rangle$ \\
81-85 & $|\!\uparrow_{ 1}, \uparrow_{ 1}, \downarrow_{ 3} , \downarrow_{ 4} \rangle$ &  $|\!\uparrow_{ 1}, \uparrow_{ 2} , \downarrow_{ 3} , \downarrow_{ 4} \rangle$ &
$|\!\uparrow_{ 1}, \downarrow_{ 3} , \uparrow_{ 3} , \downarrow_{ 4} \rangle$ &  $|\!\uparrow_{ 1}, \downarrow_{ 3} , \downarrow_{ 4} , \uparrow_{ 4} \rangle$ &
$|\!\uparrow_{ 2} , \uparrow_{ 2} , \downarrow_{ 3} , \downarrow_{ 4} \rangle$ \\
86-90 & $|\!\uparrow_{ 2} , \downarrow_{ 3} , \uparrow_{ 3} , \downarrow_{ 4} \rangle$ &
$|\!\uparrow_{ 2} , \downarrow_{ 3} , \downarrow_{ 4} , \uparrow_{ 4} \rangle$ &  $|\!\downarrow_{ 3} , \uparrow_{ 3} , \uparrow_{ 3} , \downarrow_{ 4} \rangle$ &
$|\!\downarrow_{ 3} , \uparrow_{ 3} , \downarrow_{ 4} , \uparrow_{ 4} \rangle$ &  $|\!\downarrow_{ 3} , \downarrow_{ 4} , \uparrow_{ 4} , \uparrow_{ 4} \rangle$ \\
91-95 & $|\!\uparrow_{ 1}, \uparrow_{ 1}, \downarrow_{ 4} , \downarrow_{ 4} \rangle$ &  $|\!\uparrow_{ 1}, \uparrow_{ 2} , \downarrow_{ 4} , \downarrow_{ 4} \rangle$ &
$|\!\uparrow_{ 1}, \uparrow_{ 3} , \downarrow_{ 4} , \downarrow_{ 4} \rangle$ &  $|\!\uparrow_{ 1}, \downarrow_{ 4} , \downarrow_{ 4} , \uparrow_{ 4} \rangle$ &
$|\!\uparrow_{ 2} , \uparrow_{ 2} , \downarrow_{ 4} , \downarrow_{ 4} \rangle$ \\
96-100 & $|\!\uparrow_{ 2} , \uparrow_{ 3} , \downarrow_{ 4} , \downarrow_{ 4} \rangle$ &
$|\!\uparrow_{ 2} , \downarrow_{ 4} , \downarrow_{ 4} , \uparrow_{ 4} \rangle$ &  $|\!\uparrow_{ 3} , \uparrow_{ 3} , \downarrow_{ 4} , \downarrow_{ 4} \rangle$ &
$|\!\uparrow_{ 3} , \downarrow_{ 4} , \downarrow_{ 4} , \uparrow_{ 4} \rangle$ &  $|\!\downarrow_{ 4} , \downarrow_{ 4} , \uparrow_{ 4} , \uparrow_{ 4} \rangle$\\
\hline
\end{tabular}
\caption{\label{STbl:workingbasis}Working basis $\mathcal{H}$.}
\end{table}

\subsection*{Analysis of the g-BHM}
The four-body ring-exchange process can be observed only in the plaquette initially filled with four atoms. Considering our protocol of state initialization, we define a working basis of \hbox{$\mathcal{H}\!=\! \{|\!\uparrow_i,\uparrow_j,\downarrow_k,\downarrow_l\rangle\}$}  as shown in Table.\ref{STbl:workingbasis}, for modeling the dynamical evolutions. Here the commas separate the occupations of the four spins, the subscripts \hbox{$i,j,k,l \!\in\! \{1,2,3,4\}$} denote the occupied sites of the plaquette, and the dimension of  $\mathcal{H}$ is 100. Obviously, the states \hbox{$|\psi_{27}\rangle\!=\! |\!\downarrow_1,\uparrow_2,\downarrow_3,\uparrow_4\rangle$} and \hbox{$|\psi_{63}\rangle\!=\! |\!\uparrow_1,\downarrow_2,\uparrow_3,\downarrow_4\rangle$} present in $\mathcal{H}$ are equivalent to the two states in $\mathbb{H}$ as we described before.
The g-BHM Hamiltonian in this basis can be diagonalized,
therefore we obtain the energy levels of $E_j$ and the corresponding eigenstates $|\phi_j\rangle$, $j=1,2,3,\ldots,100$.

\begin{figure}[t]
\centerline{\includegraphics[width=14.5cm]{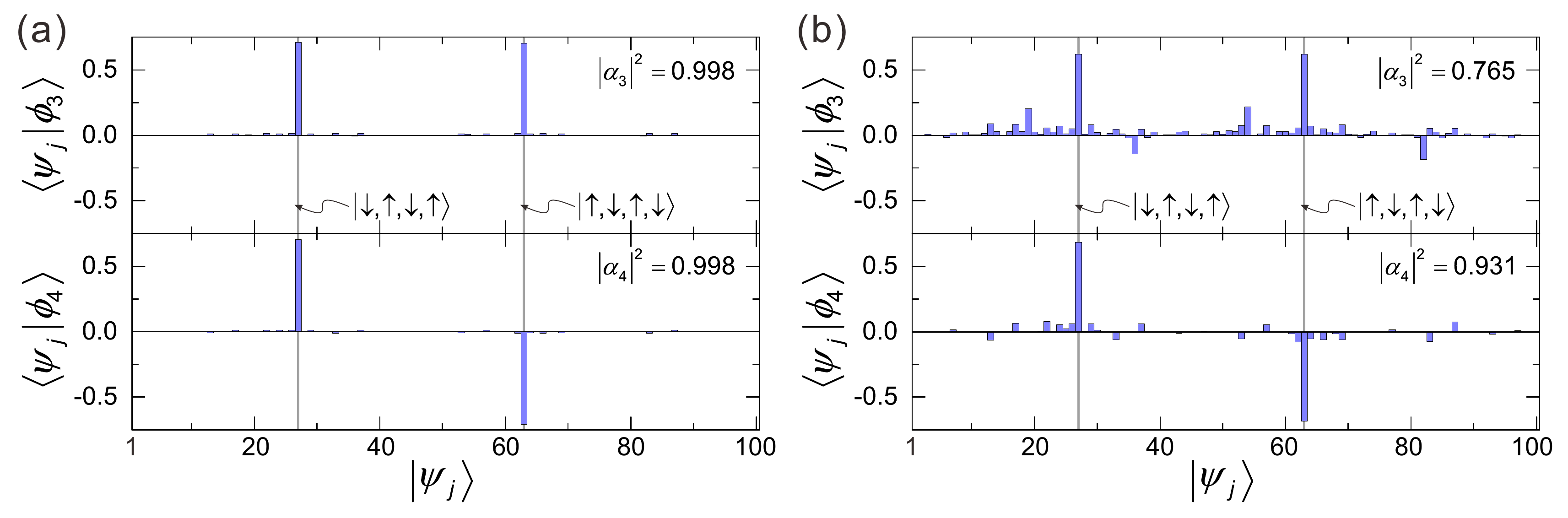}}%
\caption{\label{SFig:eigenstates}Eigenstates $|\phi_3\rangle$ and $|\phi_4\rangle$ of the g-BHM Hamiltonian. Two typical settings are shown for the experimental parameters of \textbf{(a)} $V_\mathrm{xs}\!=\!V_\mathrm{ys}\!=\!25$ $E_\mathrm{r}$, $V_\mathrm{xl}\!=\!V_\mathrm{yl}\!=\!10$ $E_\mathrm{r}$, $J_\mathrm{x}/U\!=\!J_\mathrm{y}/U\!=\!0.016$, the two eigenstates are almost equal to $|A^{\sss +}\rangle$ and $|A^{\sss -}\rangle$; \textbf{(b)} $V_\mathrm{xs}\!=\!18.2$ $E_\mathrm{r}$, $V_\mathrm{ys}\!=\!17.2$ $E_\mathrm{r}$, $V_\mathrm{xl}\!=\!V_\mathrm{yl}\!=\!10$ $E_\mathrm{r}$, $J_\mathrm{x}/U\!=\!0.082$, $J_\mathrm{y}/U\!=\!0.10$, components not belong to $\mathbb{H}$ appear in the two eigenstates.}
\end{figure}

In the regime of our experimental settings, the third and forth eigenstates take the form of
\begin{align}
|\phi_3\rangle =& \  \alpha_3|A^{\sss +}\rangle +\varepsilon_3|\phi_3^\prime\rangle,\\
|\phi_4\rangle = &\  \alpha_4|A^{\sss -}\rangle +\varepsilon_4|\phi_4^\prime\rangle,
\end{align}
where $\varepsilon_3$ and $\varepsilon_4$ denote the fractions of the components beyond the subspace of $\mathbb{H}$. Shown in Fig.\ref{SFig:eigenstates}, two groups of $\{|\phi_3\rangle,|\phi_4\rangle\}$ are derived under two different settings of the Hubbard parameters: under the condition of $J_\mathrm{x(y)}\!\rightarrow\!0$, $\{|\phi_3\rangle,|\phi_4\rangle\}$ are equivalent to $\{|A^{\sss +}\rangle,|A^{\sss -}\rangle\}$ (Fig.\ref{SFig:eigenstates}a), i.e. $\alpha_3 \!\approx\! \alpha_4 \!\approx\! 1$, $\varepsilon_3 \!\approx\! \varepsilon_4 \!\approx\! 0$; while with finite couplings (Fig.\ref{SFig:eigenstates}b), other spin components appear and the fractions of  $\{|A^{\sss +}\rangle,|A^{\sss -}\rangle\}$ are reduced, as $|\alpha_{3}|^2$ and $|\alpha_{4}|^2$ become smaller.

In the limit of strong interactions, one may expect that the ring-exchange oscillations will have a full amplitude as the superexchange processes are well suppressed. However, the ring-exchange interaction $J_{\sss \square}\!\propto\! J^4/U^3$ might be too weak as well, leading to a long period of the ring-exchange driven oscillations. The period may be beyond the coherent time of the system (about several hundreds of milliseconds), then the oscillations can not be observed in experiment. Nevertheless, in an experimentally accessible region ($J/U$ from $0.05$ to $0.15$), where $J_{\sss \square}$ is on the order of $10$ Hz, one can still prepare the system into the ring-exchange dominating regime, where the eigenstates of $\{|\phi_3\rangle,|\phi_4\rangle\}$ can be well mapped to $\{|A^{\sss +}\rangle,|A^{\sss -}\rangle\}$.

\begin{figure}[t]
\centerline{\includegraphics[width=9.87cm]{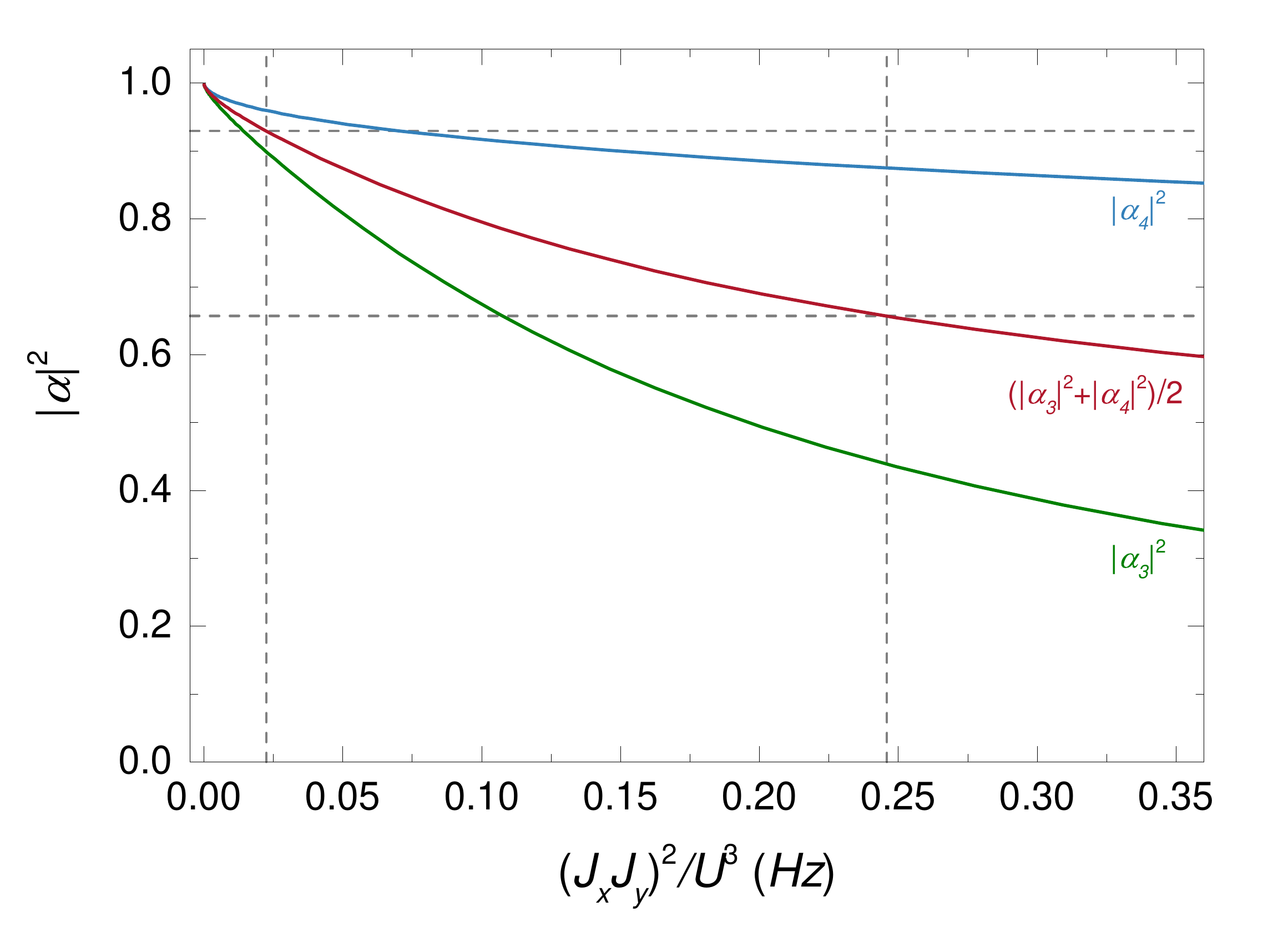}}%
\caption{\label{SFig:fidelity}Fractions of $\{|A^{\sss +}\rangle,|A^{\sss -}\rangle\}$ in  the eigensates of $\{|\phi_3\rangle,|\phi_4\rangle\}$ at different coupling parameters (green and blue lines). The average of the two projections (red line) represents the oscillation amplitude of the four-atom signals. The experimental working range is outlined by the dashed gray lines, in which the averaged projection is higher than $0.6$.}
\end{figure}

As shown in Fig.\ref{SFig:fidelity}, the projections of the eigenstates $|\alpha_3|^2$ and $|\alpha_4|^2$ are derived from the g-BHM theory with varying experimental parameters. The amplitude of the four-spin ring-exchange oscillations can be described by the averaged projections $(|\alpha_3|^2+|\alpha_4|^2)/2$. In the experimental settings of this work, the oscillation amplitude is larger than 0.6 (the full amplitude here is the four-atom signals, which can not be confused with the total atoms in main text). Therefore, in the subspace of $\mathbb {H}$, the dominating dynamics are the ring-exchange interactions. The energy gap between the two eigenstates $\Delta E \!=\! E_4 \!-\! E_3$ defines the effective ring-exchange coupling strength, which drives the dynamical evolutions between the state $|\!\downarrow,\uparrow,\downarrow,\uparrow\rangle$ and $|\!\uparrow,\downarrow,\uparrow,\downarrow\rangle$.

\subsection*{Dynamics of the noise signals}
The imperfections of the Mott insulator lead to the main source of the noise in the ring-exchange dynamics. Considering the filling properties of the Mott insulator as well as the state initializing process, it is reasonable to limit the cases of the plaquette fillings lower than 6 atoms, with spin states shown in Table.\ref{STbl:initialstates}. The most probably filled case is the plaquette with $3$ atoms, which is with one vacancy filling in the plaquette. The target state \hbox{$|\!\downarrow_1,\uparrow_2, \downarrow_3,\uparrow_4\rangle$}, which will contribute the signals in the ring-exchange evolutions, is one of the major components, with a probability of \hbox{$p_1^4/(p_1+2p_2)\!\approx\!37\%$}. All these possible initial states result a plaquette filling of $\sum p_{\mathrm{\sss ini}}n_{ \mathrm{\sss P}}\!=\!3.06$, which is almost equal to the total filling of $4(p_1+2p_2)\!=\!3.08$, the rest are the higher fillings with neglectable probabilities.

\begin{figure}[h]
\centerline{\includegraphics[width=9.87cm]{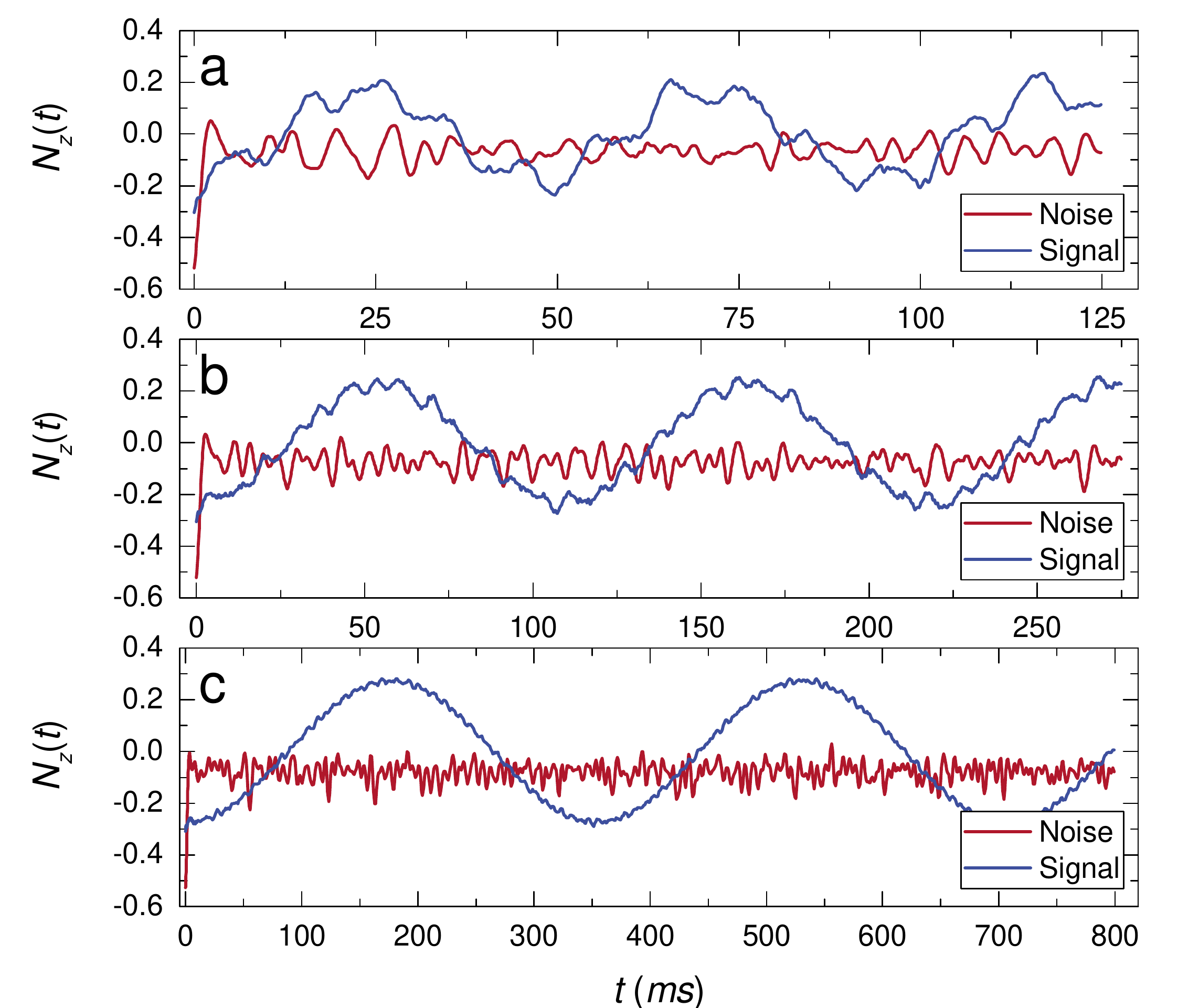}}%
\caption{\label{SFig:noise}Simulations of the dynamical evolutions of the noise (red lines) and the signal (blue lines) components. Three different simulated traces are shown under the settings of \textbf{(a)} $V_\mathrm{xs}\!=\!16.5$ $E_\mathrm{r}$, $V_\mathrm{ys}\!=\!17.2$ $E_\mathrm{r}$, $J_\mathrm{x}/U\!=\!0.12$, $J_\mathrm{y}/U\!=\!0.11$; \textbf{(b)} $V_\mathrm{xs}\!=\!18.2$ $E_\mathrm{r}$, $V_\mathrm{ys}\!=\!17.2$ $E_\mathrm{r}$, $J_\mathrm{x}/U\!=\!0.082$, $J_\mathrm{y}/U\!=\!0.10$; \textbf{(c)} $V_\mathrm{xs}\!=\!19.2$ $E_\mathrm{r}$, $V_\mathrm{ys}\!=\!18.2$ $E_\mathrm{r}$, $J_\mathrm{x}/U\!=\!0.064$, $J_\mathrm{y}/U\!=\!0.075$. }
\end{figure}

Including these initial components into the g-BHM simulations with the corresponding filling probabilities and the efficiencies of the MW pulses, it leads to a fast decay process emerging in the beginning of the evolutions, which well matches with the experiment (see Fig.\ref{SFig:noise} and the insets of Fig.2 in the main text). After the fast decay, the noise components together will become a fast fluctuating signal around a steady value, while the signal state \hbox{$|\!\downarrow_1,\uparrow_2, \downarrow_3,\uparrow_4\rangle$} will evolve under the ring-exchange interactions with a slower frequency (Fig.\ref{SFig:noise}).
Although the simulated traces are the coherent evolutions of the g-BHM Hamiltonians, which do not include the interactions with the environment and the decay mechanisms, it gives an instructive picture of the dynamical evolutions.
Therefore, the four-body ring-exchange oscillations can be time resolved in our experiment.
\begin{table}[h]
\centering
\begin{tabular}{|c|c|c|}
  \hline
  $n_{ \mathrm{\sss P}} $ & $p_\mathrm{\sss ini}$ & $|\psi_{\mathrm{\sss ini}}\rangle$\\
  \hline
  1 & $p_0^3 p_1\!=\!1.14\%$  &$|\!\downarrow_1\rangle$, $|\!\uparrow_2\rangle$, $|\!\downarrow_3\rangle$, $|\!\uparrow_4\rangle$\\
  \hline
  \multirow{2}{*}{2} & $p_0^3 p_2 \!=\! 0.03\%$ & $|\!\downarrow_1,\downarrow_1\rangle$,  $|\!\uparrow_2,\uparrow_2\rangle$, $|\!\downarrow_3,\downarrow_3\rangle$, $|\!\uparrow_4,\uparrow_4\rangle$ \\ \cline{2-3}
  &$ p_0^2 p_1^2\!=\!3.33\%$ & $|\!\downarrow_1,\uparrow_2\rangle$, $|\!\downarrow_1,\downarrow_3\rangle$, $|\!\downarrow_1,\uparrow_4\rangle$,  $|\!\uparrow_2,\downarrow_3\rangle$, $|\!\uparrow_2,\uparrow_4\rangle$, $|\!\downarrow_3,\uparrow_4\rangle$\\
  \hline
  \multirow{5}{*}{3} &$p_0 p_1^3\!=\!9.73\%$ & $|\!\downarrow_1,\uparrow_2,\uparrow_4\rangle$, $|\!\uparrow_2,\downarrow_3,\uparrow_4\rangle$, $|\!\downarrow_1,\uparrow_2,\downarrow_3\rangle$, $|\!\downarrow_1,\downarrow_3,\uparrow_4\rangle$\\ \cline{2-3}
  & \multirow{4}{*}{$ p_0^2 p_1 p_2\!=\!0.37\%$}& $|\!\downarrow_1,\downarrow_1,\uparrow_2\rangle$, $|\!\downarrow_1,\downarrow_1,\downarrow_3\rangle$, $|\!\downarrow_1,\downarrow_1,\uparrow_4\rangle$,\\
  & & $|\!\downarrow_1,\uparrow_2,\uparrow_2\rangle$, $|\!\uparrow_2,\uparrow_2,\downarrow_3\rangle$, $|\!\uparrow_2,\uparrow_2,\uparrow_4\rangle$,\\
  & & $|\!\downarrow_1,\downarrow_3,\downarrow_3\rangle$, $|\!\uparrow_2,\downarrow_3,\downarrow_3\rangle$, $|\!\downarrow_3,\downarrow_3,\uparrow_4\rangle$,\\
  & & $|\!\downarrow_1,\uparrow_4,\uparrow_4\rangle$, $|\!\uparrow_2,\uparrow_4,\uparrow_4\rangle$,  $|\!\downarrow_3,\uparrow_4,\uparrow_4\rangle$\\
  \hline
  \multirow{7}{*}{4} & \textcolor{red}{$p_1^4\!=\!28.40\%$}  &  $|\!\downarrow_1,\uparrow_2,\downarrow_3,\uparrow_4\rangle$\\ \cline{2-3}
  & \multirow{4}{*}{$ p_0 p_1^2 p_2\!=\!0.27\%$} & $|\!\downarrow_1,\downarrow_1,\uparrow_2,\downarrow_3\rangle$, $|\!\downarrow_1,\downarrow_1,\uparrow_2,\uparrow_4\rangle$, $|\!\downarrow_1,\downarrow_1,\downarrow_3,\uparrow_4\rangle$,\\
  & & $|\!\downarrow_1,\uparrow_2,\uparrow_2,\downarrow_3\rangle$, $|\!\downarrow_1,\uparrow_2,\uparrow_2,\uparrow_4\rangle$, $|\!\downarrow_1,\uparrow_2,\downarrow_3,\downarrow_3\rangle$,\\
  & & $|\!\downarrow_1,\uparrow_2,\uparrow_4,\uparrow_4\rangle$, $|\!\downarrow_1,\downarrow_3,\downarrow_3,\uparrow_4\rangle$, $|\!\downarrow_1,\downarrow_3,\uparrow_4,\uparrow_4\rangle$,\\
  & & $|\!\uparrow_2,\uparrow_2,\downarrow_3,\uparrow_4\rangle$, $|\!\uparrow_2,\downarrow_3,\downarrow_3,\uparrow_4\rangle$, $|\!\uparrow_2,\downarrow_3,\uparrow_4,\uparrow_4\rangle$\\ \cline{2-3}
  & \multirow{2}{*}{$p_0^2 p_2^2\!\approx\!0$} & $|\!\downarrow_1,\downarrow_1,\uparrow_2,\uparrow_2\rangle$, $|\!\downarrow_1,\downarrow_1,\uparrow_4,\uparrow_4\rangle$,\\
  & &  $|\!\uparrow_2,\uparrow_2,\downarrow_3,\downarrow_3\rangle$, $|\!\downarrow_3,\downarrow_3,\uparrow_4,\uparrow_4\rangle$ \\
  \hline
  \multirow{6}{*}{5} & \multirow{4}{*}{$p_0 p_1 p_2^2\!=\!0.01\%$} & $|\!\downarrow_1,\downarrow_1,\uparrow_2,\uparrow_2,\downarrow_3\rangle$,   $|\!\downarrow_1,\downarrow_1,\uparrow_2,\uparrow_2,\uparrow_4\rangle$,\\
  & & $|\!\downarrow_1,\downarrow_1,\uparrow_2,\uparrow_4,\uparrow_4\rangle$,  $|\!\downarrow_1,\downarrow_1,\downarrow_3,\uparrow_4,\uparrow_4\rangle$,\\
  & & $|\!\downarrow_1,\uparrow_2,\uparrow_2,\downarrow_3,\downarrow_3\rangle$,  $|\!\uparrow_2,\uparrow_2,\downarrow_3,\downarrow_3,\uparrow_4\rangle$,\\
  & & $|\!\downarrow_1,\downarrow_3,\downarrow_3,\uparrow_4,\uparrow_4\rangle$, $|\!\uparrow_2,\downarrow_3,\downarrow_3,\uparrow_4,\uparrow_4\rangle$\\ \cline{2-3}
  &  \multirow{2}{*}{$p_1^3 p_2\!=\!0.78\%$} & $|\!\downarrow_1,\uparrow_2,\uparrow_2,\downarrow_3,\uparrow_4\rangle$, $|\!\downarrow_1,\uparrow_2,\downarrow_3,\uparrow_4,\uparrow_4\rangle$,\\
  & & $|\!\downarrow_1,\downarrow_1,\uparrow_2,\downarrow_3,\uparrow_4\rangle$, $|\!\downarrow_1,\uparrow_2,\downarrow_3,\downarrow_3,\uparrow_4\rangle$\\
  \hline
\end{tabular}
\caption{\label{STbl:initialstates}Possible initial spin configurations in the plaquette and the corresponding probabilities. Here $p_j$ is the probability for the case of $j$ particles filled in one site, $p_\mathrm{\sss ini}$ is the probability for the case of spin configuration $|\psi_\mathrm{\sss ini}\rangle$, and $n_\mathrm{\sss P}$ denotes the total particle number in the plaquette of state $|\psi_\mathrm{\sss ini}\rangle$.}
\end{table}

\end{document}